\documentclass[aps,epsfig,prl,twocolumn,superscriptaddress,showpacs,showkeys,email]{revtex4}
\usepackage{graphics}
\begin{document} 

\title{\bf Cooperativity and the origins of rapid, single-exponential
kinetics in protein folding}

\author{Patr\'\i cia FN Fa\'\i sca }
\email{patnev@cii.fc.ul.pt}
\affiliation{Centro de F\'\i sica Te\'orica e Computacional da Universidade de Lisboa, Av. Prof. Gama Pinto 2, 1649-003 Lisboa Codex, Portugal}

\author{Kevin W. Plaxco} 
\email{kwp@chem.ucsb.edu}
\affiliation{Department of Chemistry and Biochemistry, University of California, Santa Barbara, Santa Barbara CA 93106, USA } 

\begin{abstract}
{\bf The folding of naturally occurring, single domain proteins is usually well-described as a simple, single exponential process lacking significant trapped states. Here we further explore the hypothesis that the smooth energy landscape this implies, and the rapid kinetics it engenders, arises due to the extraordinary thermodynamic cooperativity of protein folding.  Studying Miyazawa-Jernigan lattice polymers we find that, even under conditions where the folding energy landscape is relatively optimized (designed sequences folding at their temperature of maximum folding rate), the folding of protein-like heteropolymers is accelerated when their thermodynamic cooperativity enhanced by enhancing the non-additivity of their energy potentials.  At lower temperatures, where kinetic traps presumably play a more significant role in defining folding rates, we observe still greater cooperativity-induced acceleration. Consistent with these observations, we find that the folding kinetics of our computational models more closely approximate single-exponential behavior as their cooperativity approaches optimal levels. These observations suggest that the rapid folding of naturally occurring proteins is, at least in part, consequences of their remarkably cooperative folding.}
\end{abstract}

\pacs{\bf{87.15.Cc; 91.45.Ty}} \keywords{\bf{Monte Carlo simulation; optimal folding temperature, native topology, contact order, misfolded states, two-state}}

\maketitle

\section{\bf{Introduction}}

Diffusive processes, such as protein folding, progress more rapidly across a smooth energy landscape than on rough landscapes dominated by myriads of energetic barriers (Bryngelson 1995).  Exhaustive lattice and off-lattice heteropolymer simulations support this simple argument; rapid folding is observed only among those relatively rare sequences that lack well-populated, trapped intermediates (Mirny 1996; Chan and Dill 1997; Chan and Dill 1998; Cieplak 1998). Similarly supporting this argument, the rapid folding of naturally occurring, single domain proteins is almost always associated with energy landscapes that are smooth relative to $k_{B}$T, as indicated by the observation of simple, single exponential kinetics even at the lowest temperatures that can be investigated experimentally (Gillespie and Plaxco 2000; 2004). 
\par
While the relationship between smooth energy landscapes and rapid folding is well established (Bryngelson 1995; Mirny 1996; Chan and Dill 1997; Chan and Dill 1998; Cieplak 1998), the perhaps more subtle question of the molecular origins of this smoothness has seen relatively less attention.  One of the few attempts to address this issue directly is the work of Shakhnovich and co-workers, who have noted that the landscapes of the vast majority of lattice polymer sequences are very rough, and that smooth lattice polymer landscapes are comparatively uncommon (Miller 2002). Based on this observation they speculate that polypeptides exhibiting smooth landscapes are similarly uncommon and that the rapid folding of naturally occurring proteins arises as a consequence of evolutionary selections aimed at ensuring that this critical property is achieved (e.g. refs Abkevich 1996; Gutin 1998a; Mirny 1998).  Unfortunately, however, experimental studies have not supported this proposal; the folding of de novo designed proteins -proteins produced without any regard to the roughness of their energy landscapes- is generally rapid (Gillespie 2003; Zhu 2003; Scalley-Kim and Baker 2004; Khulman and Baker 2004), suggesting that smooth landscapes are a relatively common property of thermodynamically foldable polypeptides(Gillespie 2003; Gillespie and Plaxco, 2004). Thus the ultimate origins of the smooth folding energy landscapes observed for naturally occurring proteins remains unclear.
\par
It has been speculated that the discrepancy between the rapid, trap-free folding observed for most small proteins and the generally slow, trap-dominated folding of lattice and off-lattice models (Jewett 2003) may arise because, in contrast to the folding of most computational models, protein folding is thermodynamically cooperative  (Kaya and Chan 2003; Kaya and Chan 2003a; Chan 2004; Kaya 2005).  Thermodynamic cooperativity, usually identified by the calorimetric criterion of Privalov (see e.g. Makhatadze and Privalov 1995), is defined as a bimodal conformational population peaked around the native and unfolded state enthalpies and practically zero at intermediate enthalpies.  The word cooperativity is used to describe this effect because it is widely thought to arise from nonadditive (i.e. cooperative) interactions akin to those first proposed as an explanation of the cooperative oxygen binding of hemoglobin (Wyman and Allen 1959). The speculation that similar cooperativity (arising from similarly non-additive interactions) might account for the smooth landscapes almost universally observed for the folding of small proteins is based on the hypothesis that such will cooperativity accelerate folding more by destabilizing misfolded, trapped states than it decelerates folding by destabilizing productive intermediates (Jewett 2003; Eastwood and Wolynes 2001). 
\par
It has not proven possible to modulate the thermodynamic cooperativity of proteins in the laboratory, and thus it has not yet proven feasible to explore the relationship between cooperativity and folding rates in vitro.  It is possible, however, to modulate the cooperativity of computational protein models and thus simulation experiments provide a useful means of testing the hypothesized link.  Despite extensive simulations-based studies of the role of thermodynamic cooperativity in generating linear chevron behavior, topology-dependent rates and other signatures of two-state folding kinetics, however, the extent to which cooperativity acts to accelerate or decelerate folding has not been explicitly investigated in the prior literature (Fan 2001; Shimizu and Chan 2002; Kaya and Chan 2003; Jewett 2003; Kaya and Chan 2003a; Kaya and Chan 2003b; Cieplak 2004).  Here we explore this question via simulations of protein-like Miyazawa-Jernigan lattice polymers to which increasing degrees of cooperativity have been introduced.

\section{\bf{Results}}

\subsection{\it {Test sysytems}}

\begin{figure*}
{\rotatebox{270}{\resizebox{7.6cm}{5.6cm}{\includegraphics{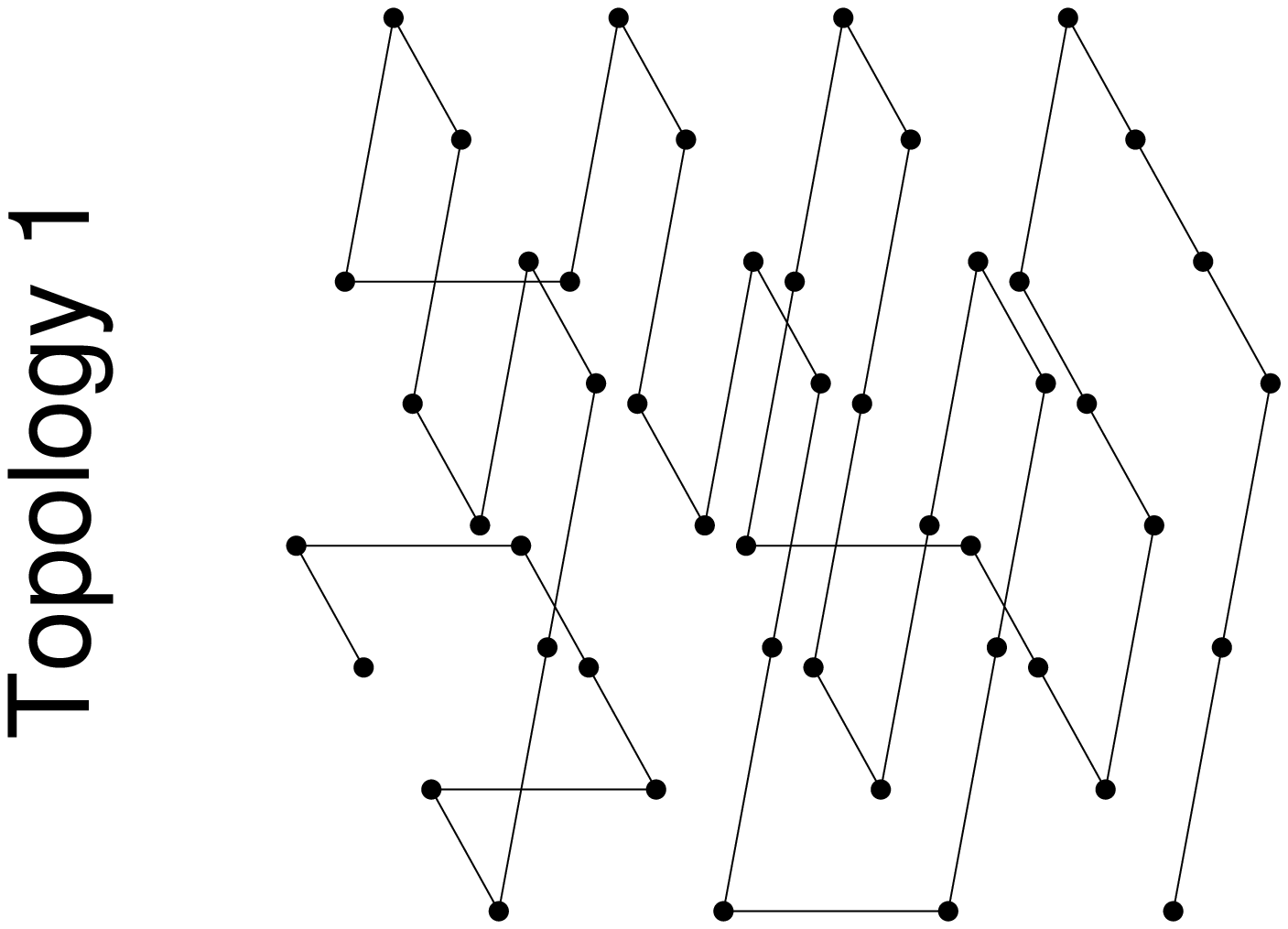}}}}
\hspace{0.5cm}
{\rotatebox{270}{\resizebox{7.6cm}{5.6cm}{\includegraphics{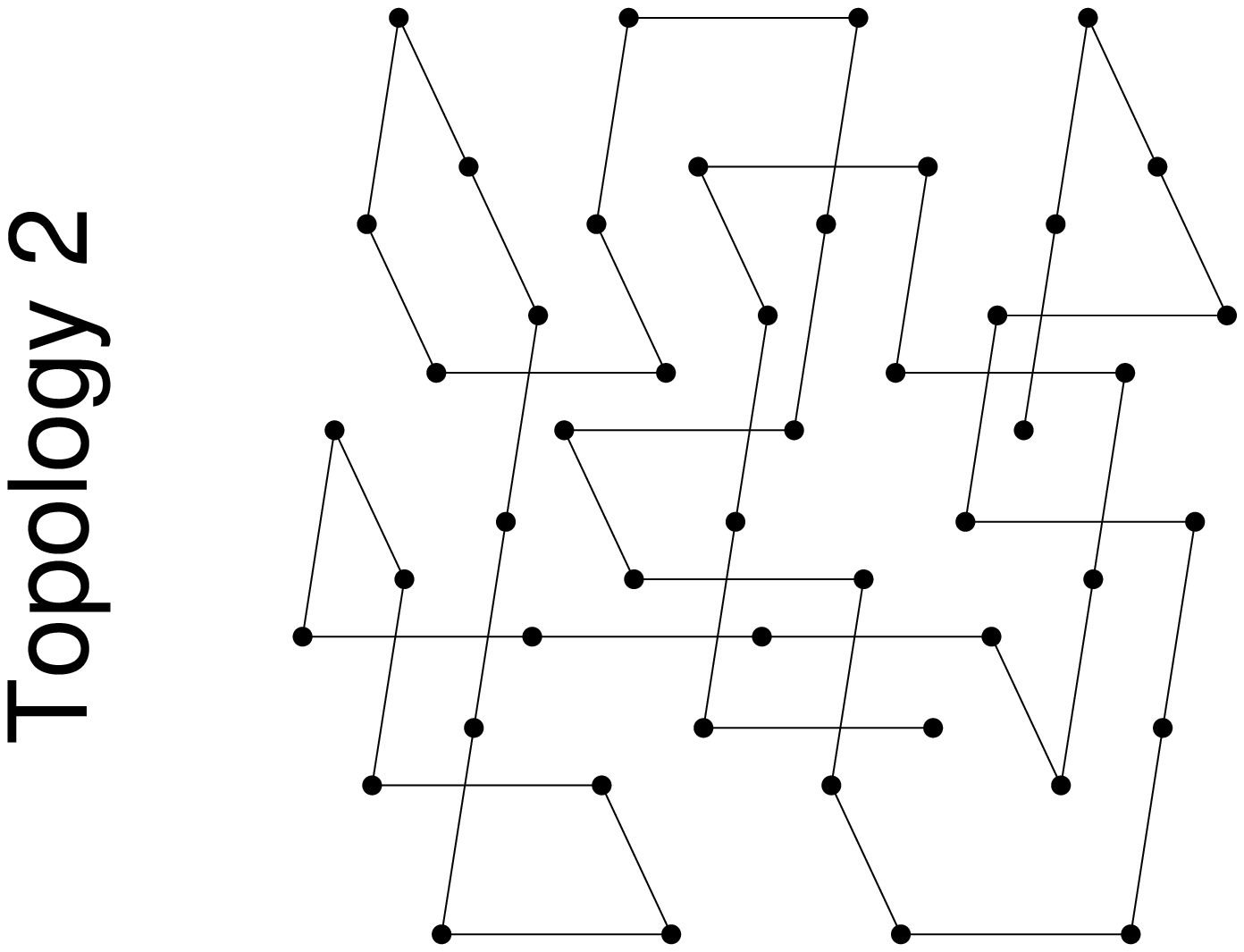}}}
\vspace{1.3cm}}
{\rotatebox{0}{\resizebox{6.6cm}{6.6cm}{\includegraphics{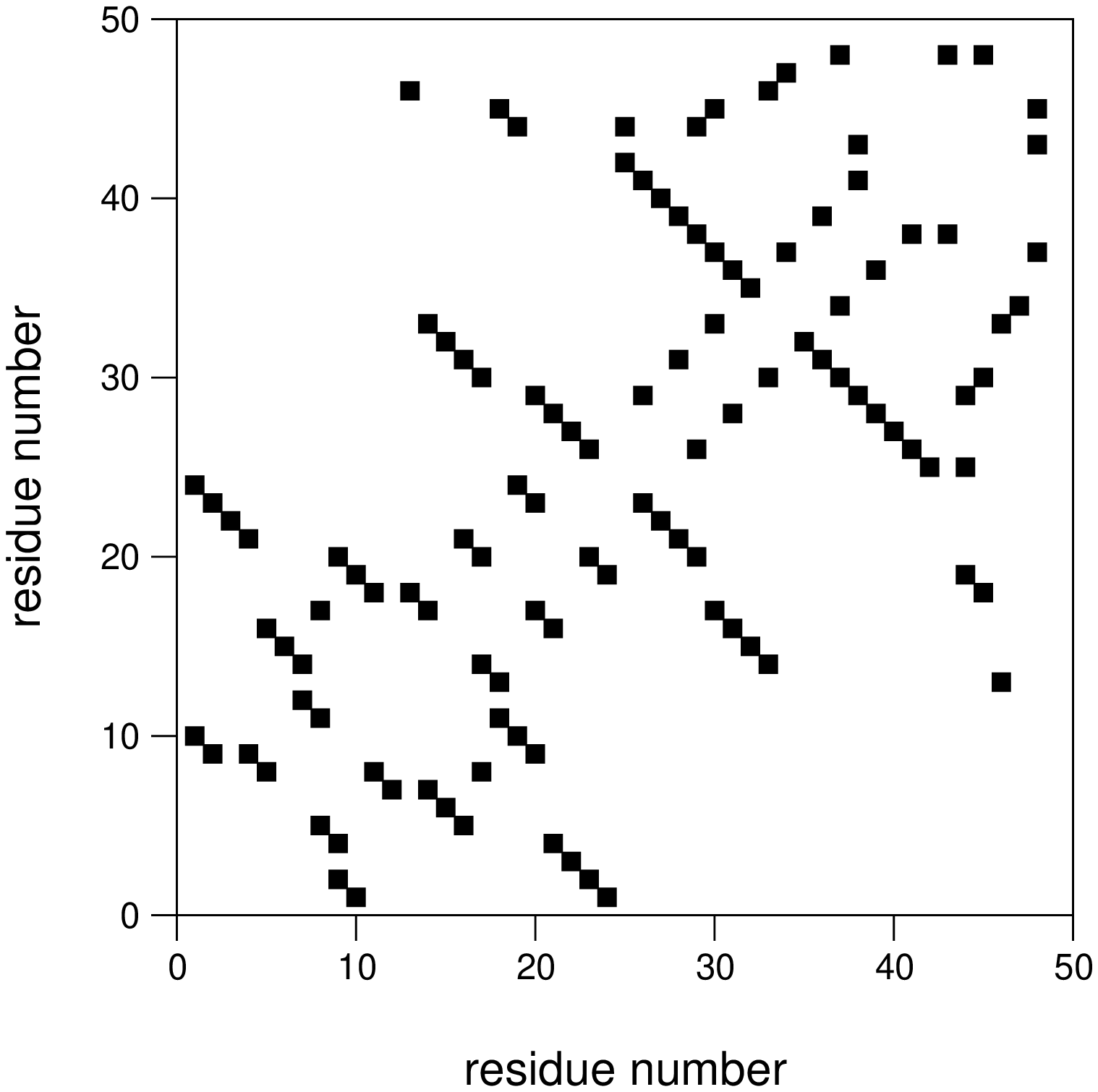}}}}
\hspace{0.5cm}
{\rotatebox{0}{\resizebox{6.6cm}{6.6cm}{\includegraphics{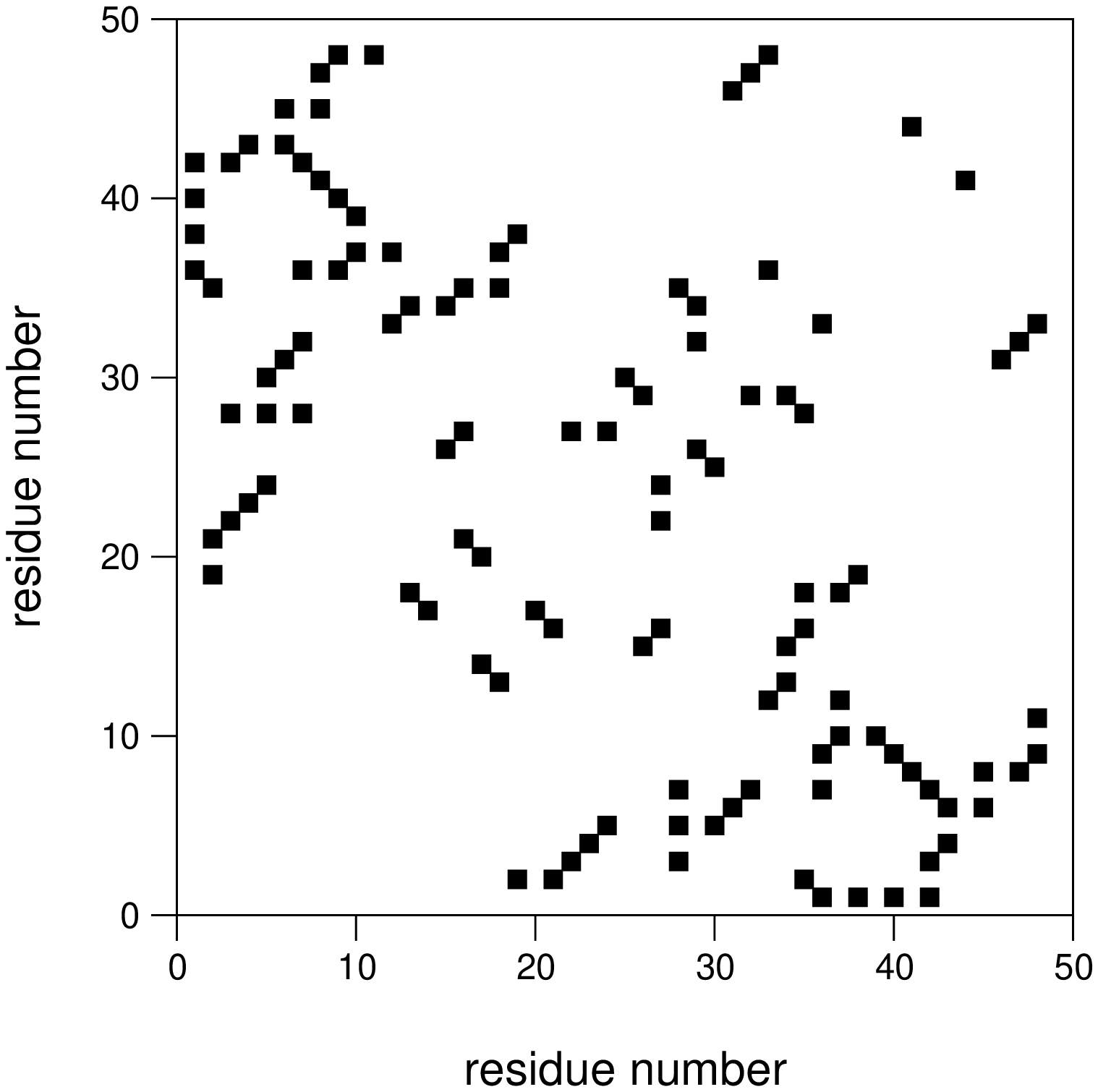}}}}
\caption{In this study we have employed sequences folding into two topologies (numbered 1 and 2), which are among the least and most complex topologies attainable for a maximally compact 48-residue lattice polymer.  (Bottom) The bands along the main diagonal in the contact maps (Saitoh 1993) of the two topologies bands indicating contacts between one amino acid and its four successors represent the structural equivalent of $\alpha$-helices.  $\beta$-sheets are represented as thick bands parallel or anti-parallel to the diagonal.}
\label{fig:no1}
\end{figure*}

We have explored the folding behavior of 48-residue lattice polymer sequences encoding two topologically distinct native structures. The two structures, named topologies 1 and 2, exhibit relative contact orders (Plaxco 1998a) of 13\% and 26\% respectively, placing them among the least and most ``complex" topologies that a maximally compact, 48-monomer structure can adopt (Figure 1; table 1). Using the design algorithm of Shakhnovich and Gutin (Shakhnovich and Gutin 1993) we obtained three, rapidly folding sequences encoding each of these distinct topologies. The sequences were designed to adopt the target topologies using the Myiazawa-Jernigan (MJ) energy parameterization,

\begin{equation}
E(\lbrace \sigma_{i} \rbrace,\lbrace \vec{r_{i}} \rbrace)=\sum_{i>j}^N
\epsilon(\sigma_{i},\sigma_{j})\Delta(\vec{r_{i}}-\vec{r_{j}}),
\label{eq:no1}
\end{equation}
\noindent
where $\lbrace \vec{r_{i}} \rbrace$ is the set of bead coordinates
that define a conformation, $\lbrace \sigma_{i} \rbrace$ represents an
amino acid sequence, and $\sigma_{i}$ stands for the chemical identity
of bead $i$.  The contact function $\Delta$ is $1$ if beads $i$ and
$j$ form a contact (that is not a covalent linkage) and is $0$
otherwise. The energy parameters $\epsilon_{i,j}$ are taken from the
$20 \times 20$ MJ matrix, derived from the distribution of contacts of
native proteins (Miyazawa and Jernigan 1985). In order to minimize the possiblity of
energy-related kinetic effects, sequences were selected that exhibit
similar native state energies, $E_{nat}$ (table 2).

\subsection{\it{Enhancing cooperativity}}

While it is clear that a remarkable feature like thermodynamic cooperativity demands highly unusual energetics, the nature of the interactions underlying the thermodynamic cooperativity of protein folding has not yet been determined (Takada 1999; Kaya and Chan 2000; Fernandez 2002; Shimizu and Chan 2002; Kaya 2005). It is thought, however, that the thermodynamic cooperativity of protein folding arises due to multi-body effects leading to nonadditive interactions (i.e. in which the formation of one ``bond" makes the formation of subsequent ``bonds" more favorable (Wyman and Allen 1959)), a hypothesis that is supported by extensive simulation studies (Eastwood and Wolynes 2001; Fernandez 2002; Kaya and Chan 2003b; Ejtehadi 2004).  In order to capture the nonadditivity that presumably underlies thermodynamic cooperativity we have employed a modified version of the MJ potential that entails many-body, nonpairwise interactions in the form of a nonlinear relationship between the energy of a conformation  $E^{'}$ and the fraction of native contacts, $Q$, it forms:

\begin{equation}
E^{'}(Q,S)=\frac{E}{Q(1-S) + S} 
\label{eq:no2}
\end{equation}

where $S$ is a parameter measuring the magnitude of the induced cooperativity and $E$ is given by Equation~\ref{eq:no1}.  Of note, $Q$ is the fraction of native contacts; when we employ a similar function that counts the total number of contacts we find that MJ polymers are not thermodynamically stable if $S$ rises even trivially above unity. Note too
that when $S$ = 1 we recover Equation 1 and the traditional, noncooperative MJ potential. As $S$ increases, however, Equation 2 promotes a steeper decrease of the protein's energy with increasing number of native contacts when folding nears completion. It does so by penalizing the energy of conformations in which only a small fraction of the total set of native contacts is formed ($Q < \approx 0.7$; data not shown) and by ``repaying" that penalty after that threshold is crossed. As expected, when $S$ is increased as per equation~\ref{eq:no2} the thermodynamic cooperativity of our model systems is significantly enhanced; both the enthalpy distribution and conformational distribution of our model polymers become strongly biomodal (i.e. see e.g. Makhatadze and Privalov 1995) as $S$ is increased from to $S_{opt}$ (Figure 2)

\begin{figure*}
{\rotatebox{270}{\resizebox{7.5cm}{7.5cm}{\includegraphics{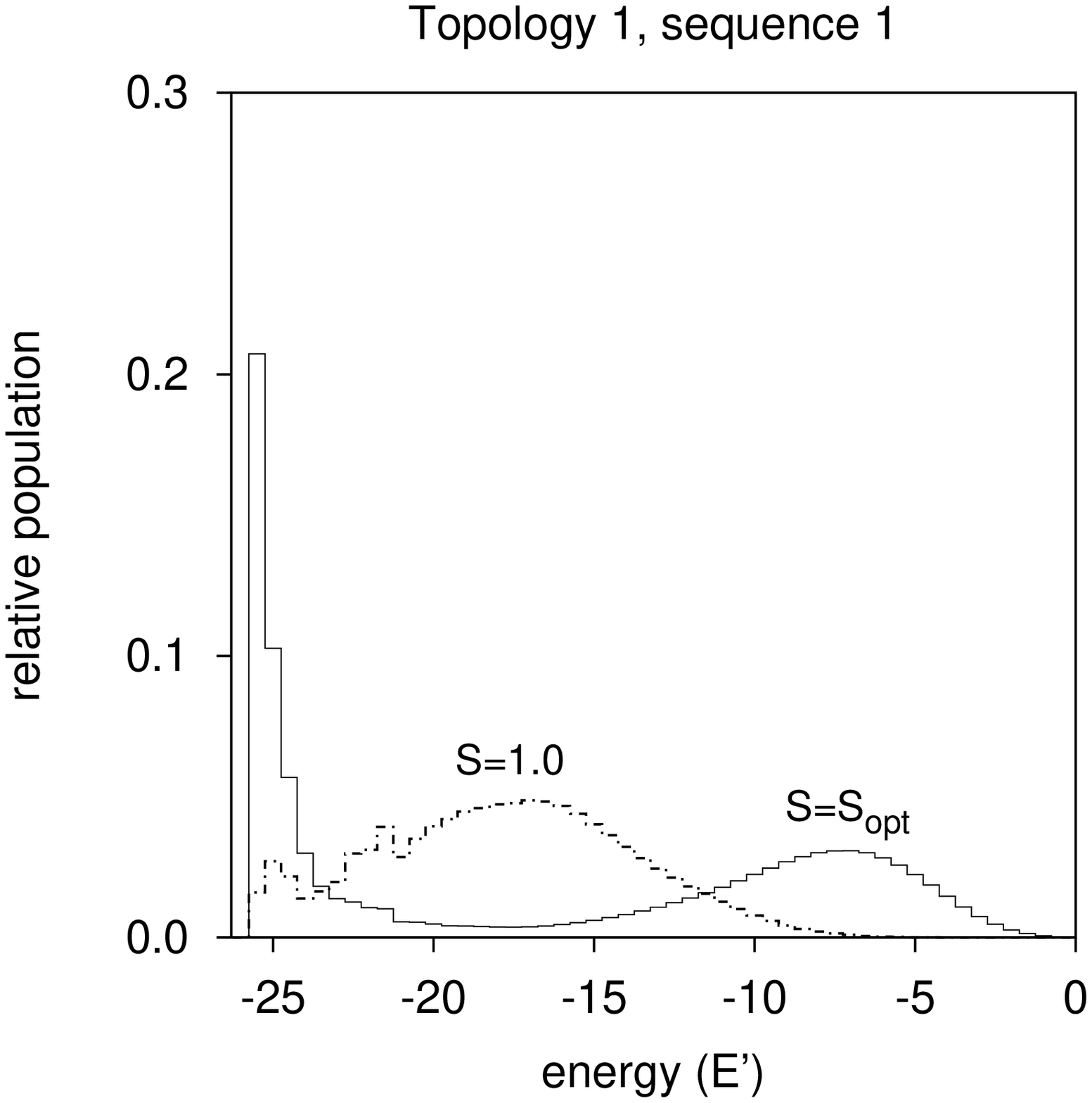}}}}
\hspace{0.5cm}
{\rotatebox{270}{\resizebox{7.5cm}{7.5cm}{\includegraphics{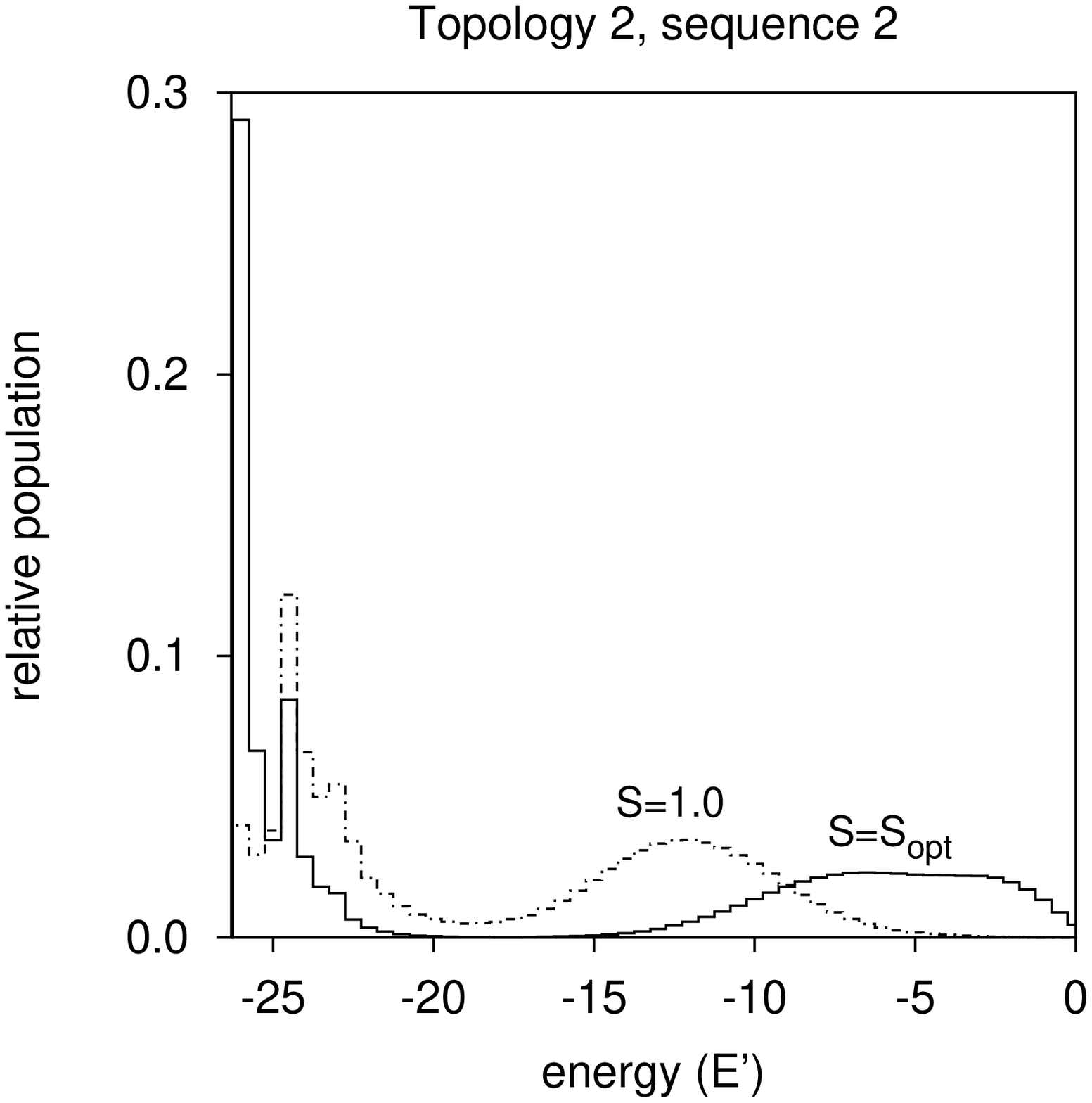}}}
\vspace{1.0cm}}
{\rotatebox{270}{\resizebox{7.5cm}{7.5cm}{\includegraphics{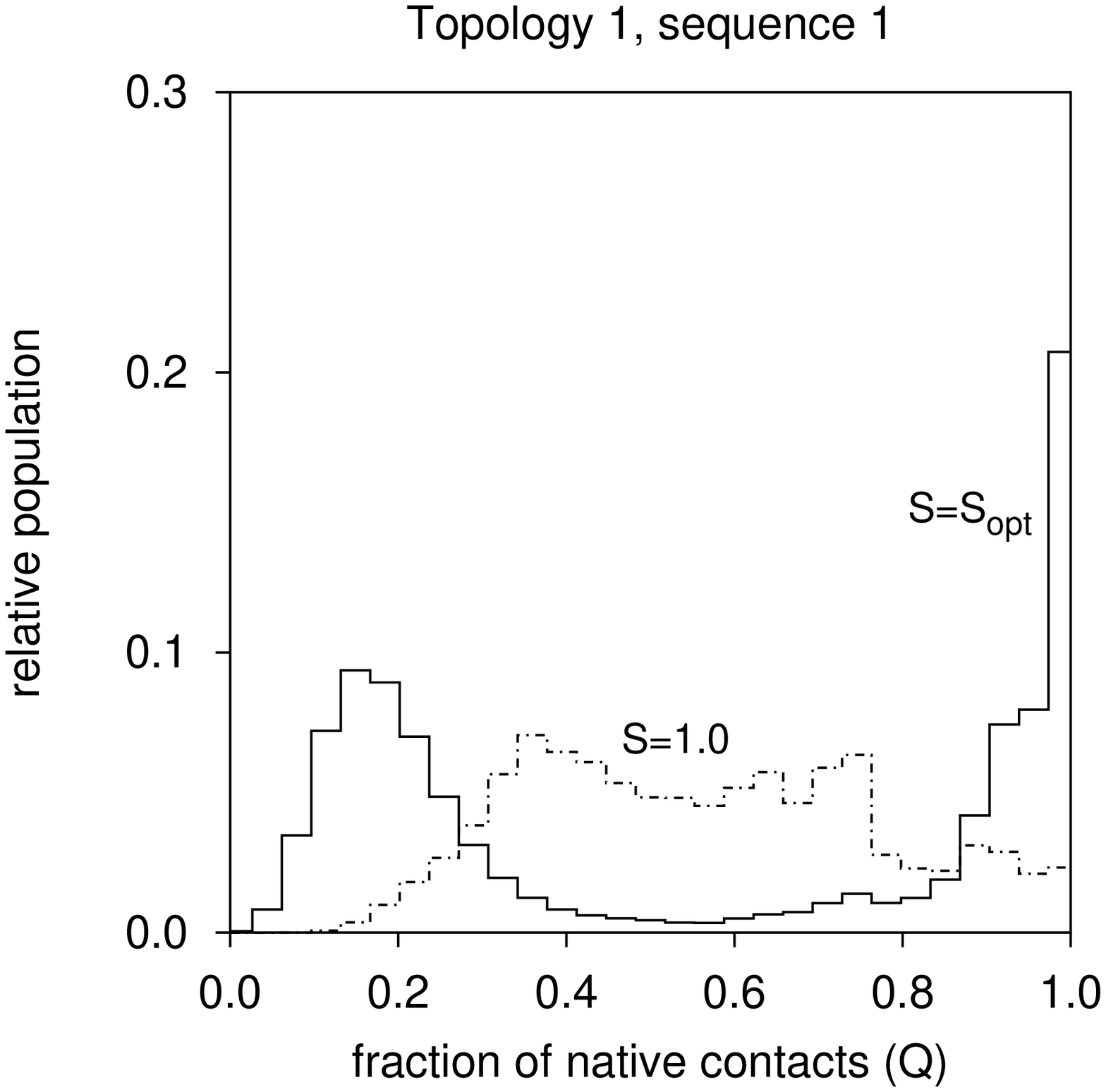}}}}
\hspace{0.5cm}
{\rotatebox{270}{\resizebox{7.5cm}{7.5cm}{\includegraphics{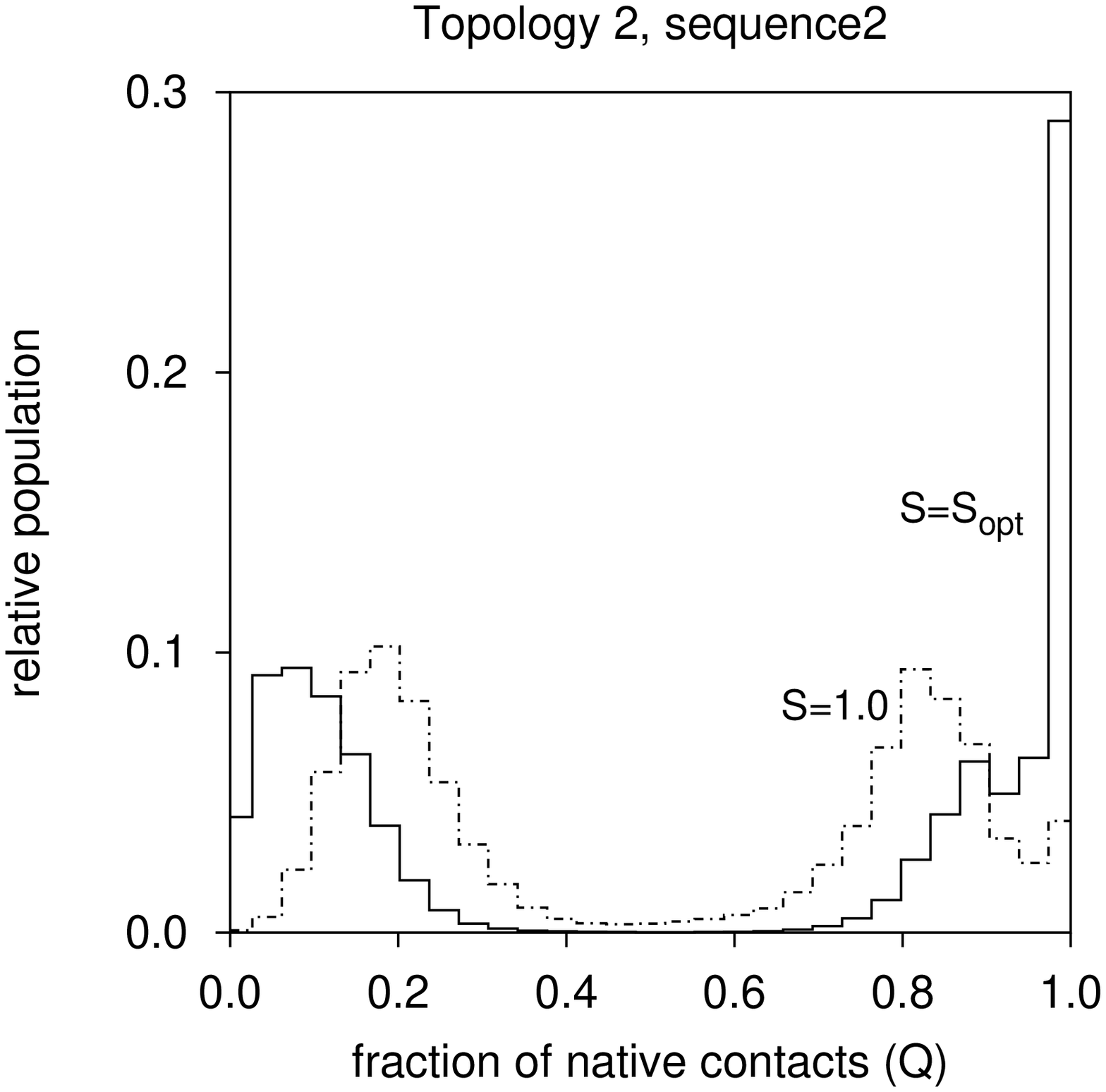}}}}
\caption{Enhancing the nonadditivity (i.e. cooperativity -see Wymann and Allen, 1959) of the MJ energy potential enhances the thermodynamic cooperativity (Makhatadze and Privalov 1995) of MJ lattice polymer folding.  As illustrated here using two representative sequences at their folding transition temperatures, T$_{f}$, the energy distributions (top row) and conformational distributions (bottom row) of our model polymers become more strongly biomodal (i.e. more thermodynamically cooperative - see e.g. Makhatadze and Privalov 1995) as $S$ is increased from 1 to $S_{opt}$. In particular, the population of fully native molecules (bottom row, $Q$ = 1.0) is enhanced significantly at higher values of $S$.}
\label{fig:no2}
\end{figure*}

\subsection{\it {The kinetic consequences of cooperativity at the optimal folding temperature}}

\begin{figure*}[!ht] 
{\rotatebox{0}{\resizebox{8.0cm}{8.0cm}{\includegraphics {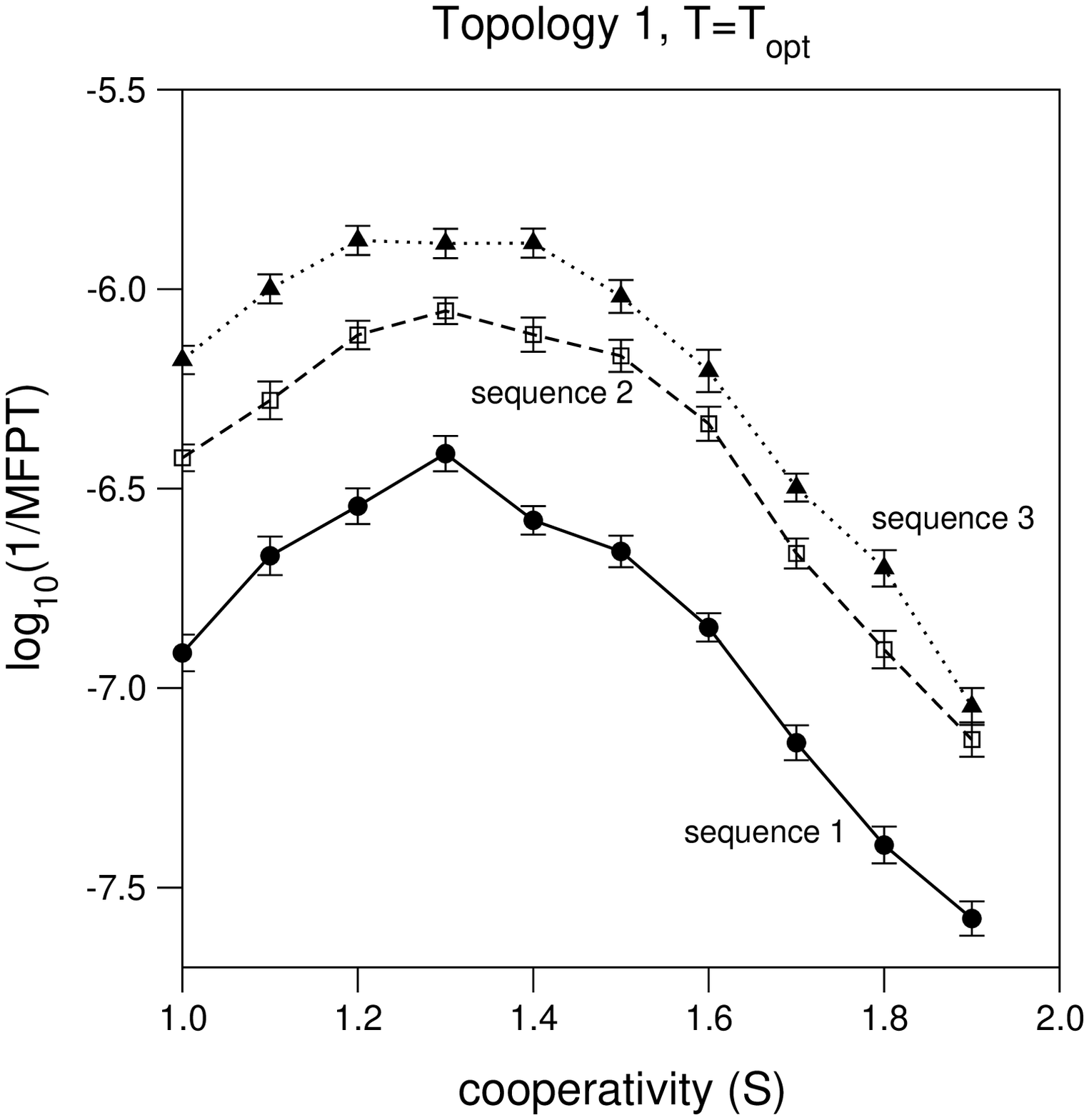}}}} \hspace{0.3cm}
{\rotatebox{0}{\resizebox{8.0cm}{8.0cm}{\includegraphics {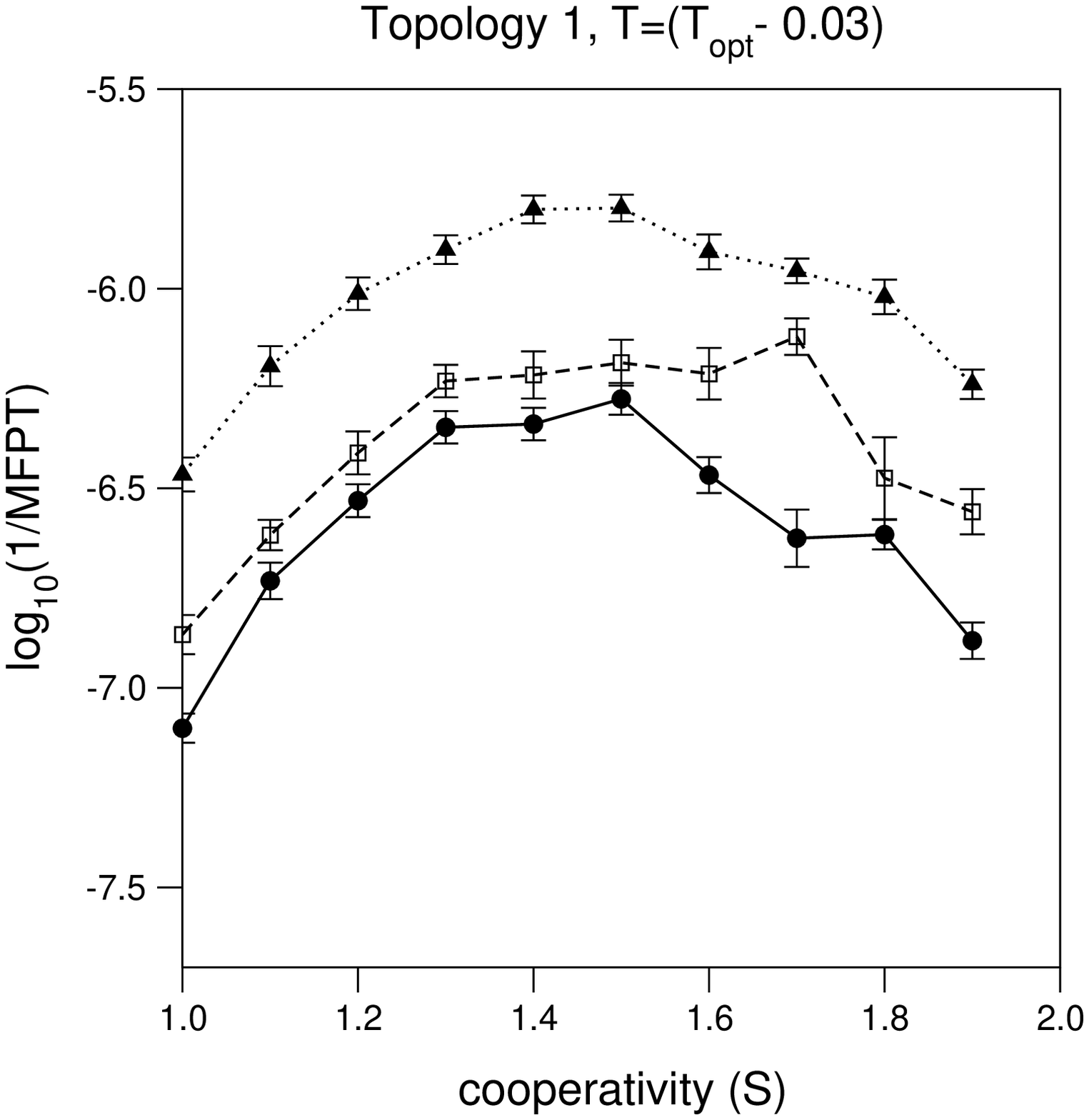}}}\vspace{1.0cm}}
{\rotatebox{0}{\resizebox{8.0cm}{8.0cm}{\includegraphics {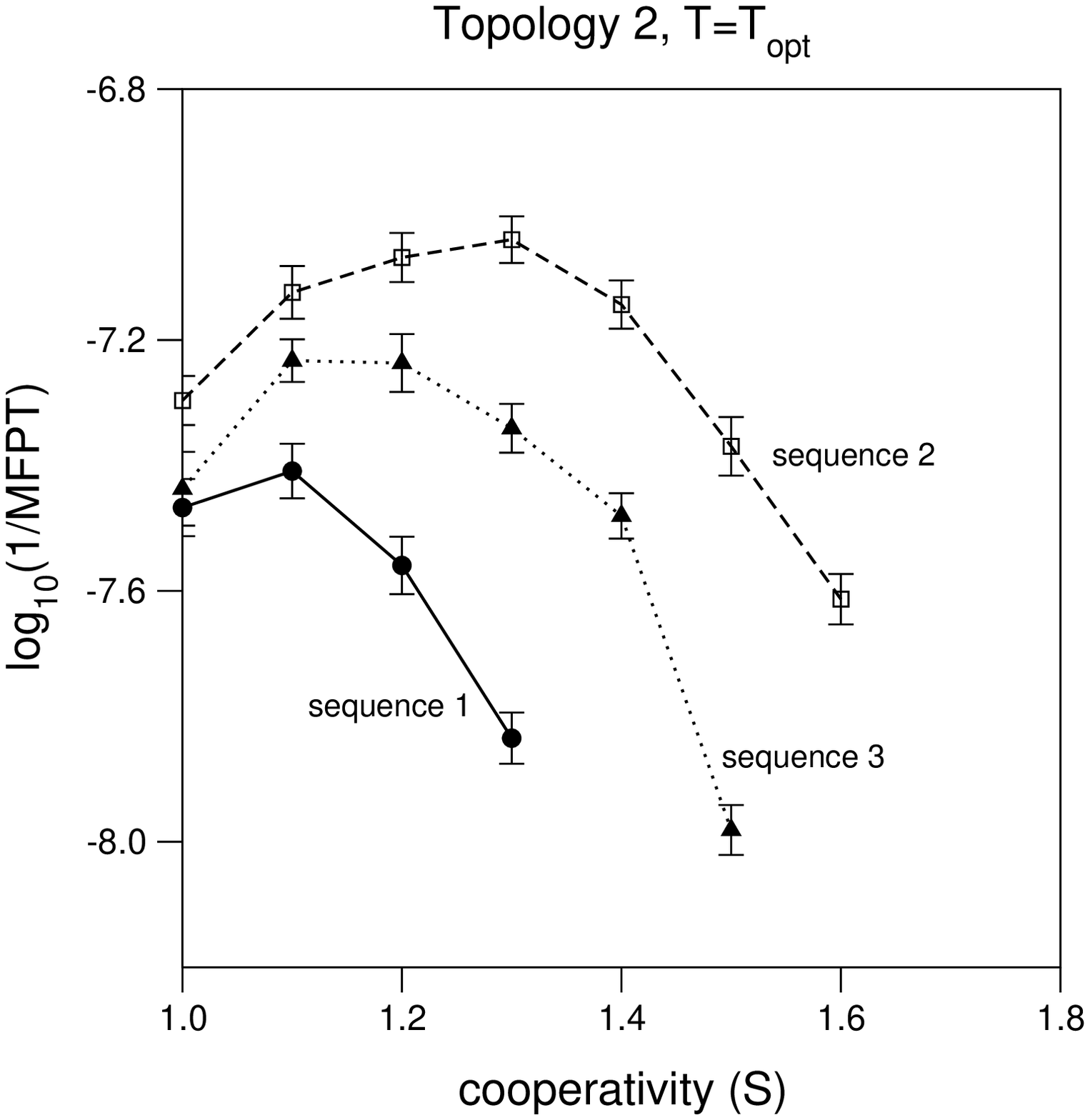}}}} \hspace{0.3cm}
{\rotatebox{0}{\resizebox{8.0cm}{8.0cm}{\includegraphics {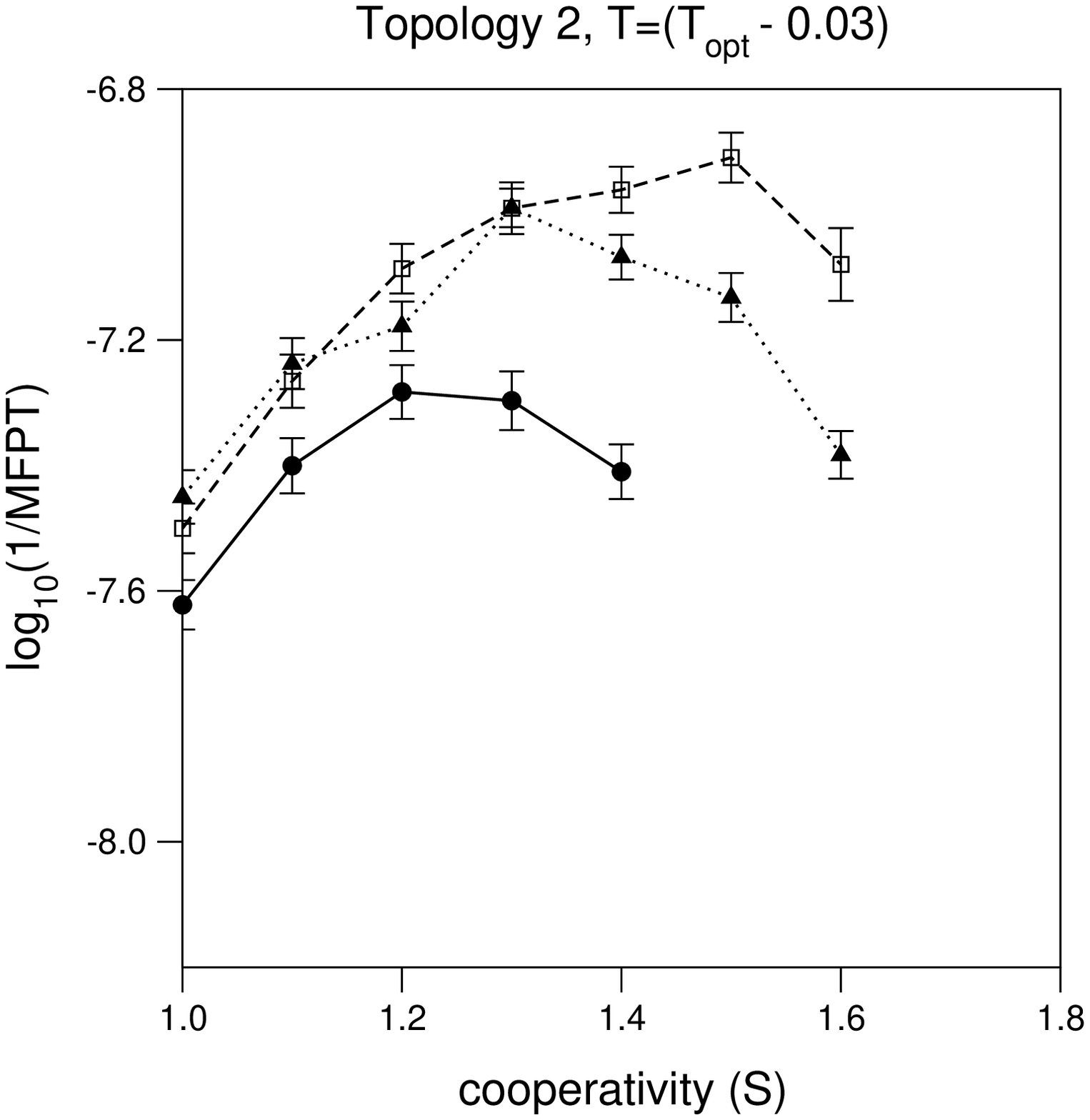}}}} \\
\caption{The folding of MJ lattice polymers is accelerated when the cooperativity of the MJ energy potential is enhanced. This effect holds for all six sequences we have investigated (adopting both low -top row- and high -bottom row- contact order structures). And while this acceleration is readily apparent even at the temperature of optimal folding (T$_{opt}$; left column), it is significantly more pronounced at lower temperatures (T = T$_{opt}$ - 0.03; right column), where kinetic traps might be expected to play a more significant role in defining kinetics. Above some optimal level of cooperativity ($S_{opt}$), however, folding decelerates with increasing cooperativity. This presumably occurs because further increases in $S$ destabilize native elements in the folding transition state, slowing folding more than the destabilization of trapped states accelerates it.}
\label{fig:no3}
\end{figure*}

In order to address the extent to which enhanced cooperativity enhances folding rates we initially performed simulations at T$_{opt}$, the temperature at which the folding rate of each sequence reaches its maximum value in the absence of added cooperativity. Under these conditions we find that folding invariably accelerates as cooperativity, $S$, increases towards its optimal level, $S_{opt}$: at their 
$S_{opt}$ the folding rates of all six of the sequences we have characterized increase by 1.14 to 3.2-fold relative to the rates observed at $S$ = 1 (Figure 3, left). Above Sopt folding rates decrease, presumably because further increases in cooperativity significantly destabilize the folding transition state.
\par
While the folding time of every test system we have investigated exhibits a similar functional dependence on $S$, we observe quantitatively different behavior among differing structures and sequences. For example, under these conditions $S_{opt}$  ranges from 1.2 to 1.3 for the three sequences that fold into topology 1 and the associated rate increases (over the non-cooperative, $S$ = 1 benchmark) range from 2.0 to 3.2-fold. In contrast, over a similar range of $S_{opt}$  we observe only 1.14 to 1.8-fold accelerations for sequences folding into the more complex topology 2.  

\par
\subsection{\it{The kinetic consequences of cooperativity at lower temperatures}}

\begin{figure}
{\rotatebox{0}{\resizebox{8.0cm}{8.0cm}{\includegraphics{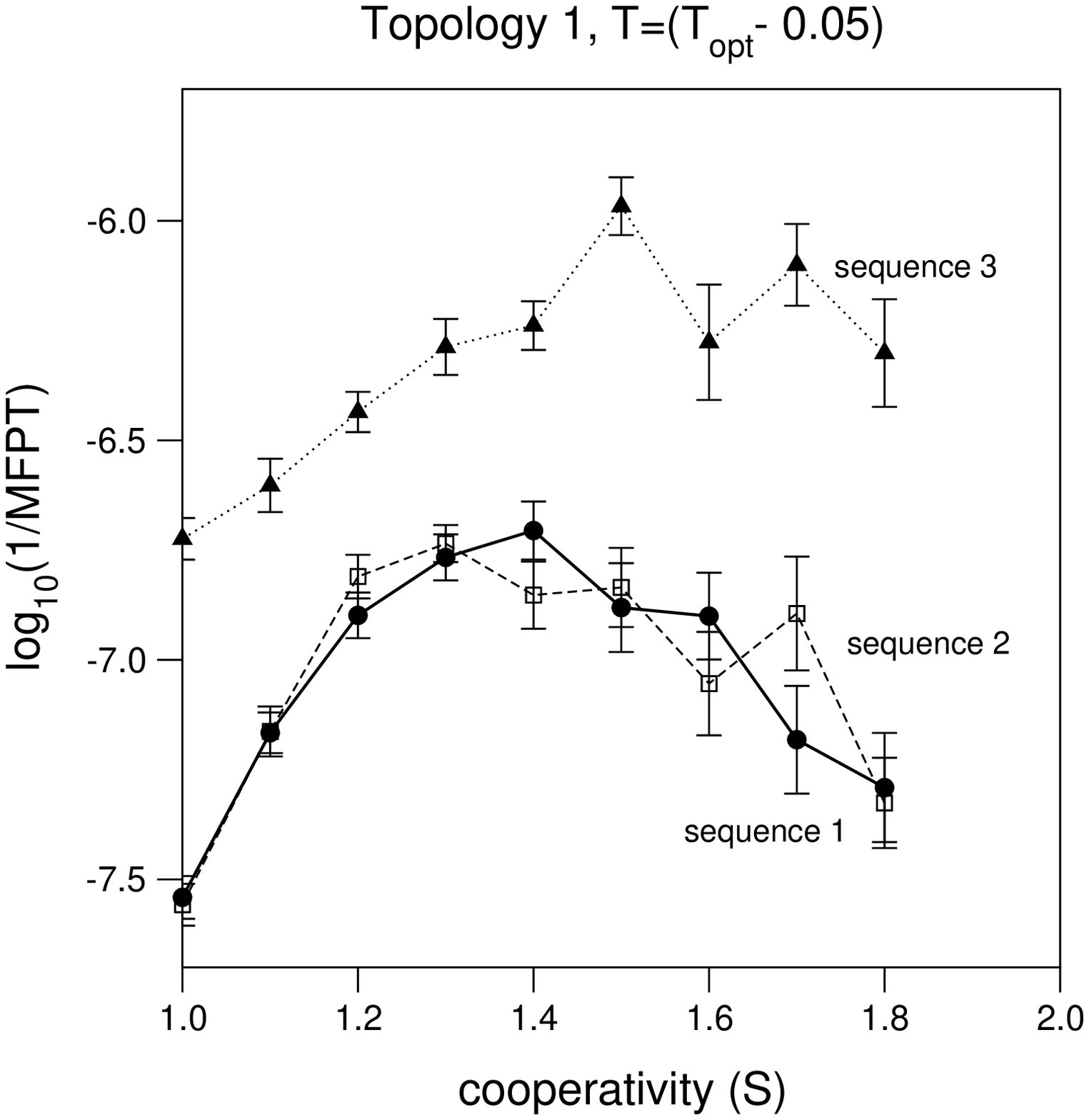}}}\vspace{1.0cm}}
{\rotatebox{0}{\resizebox{8.0cm}{8.0cm}{\includegraphics{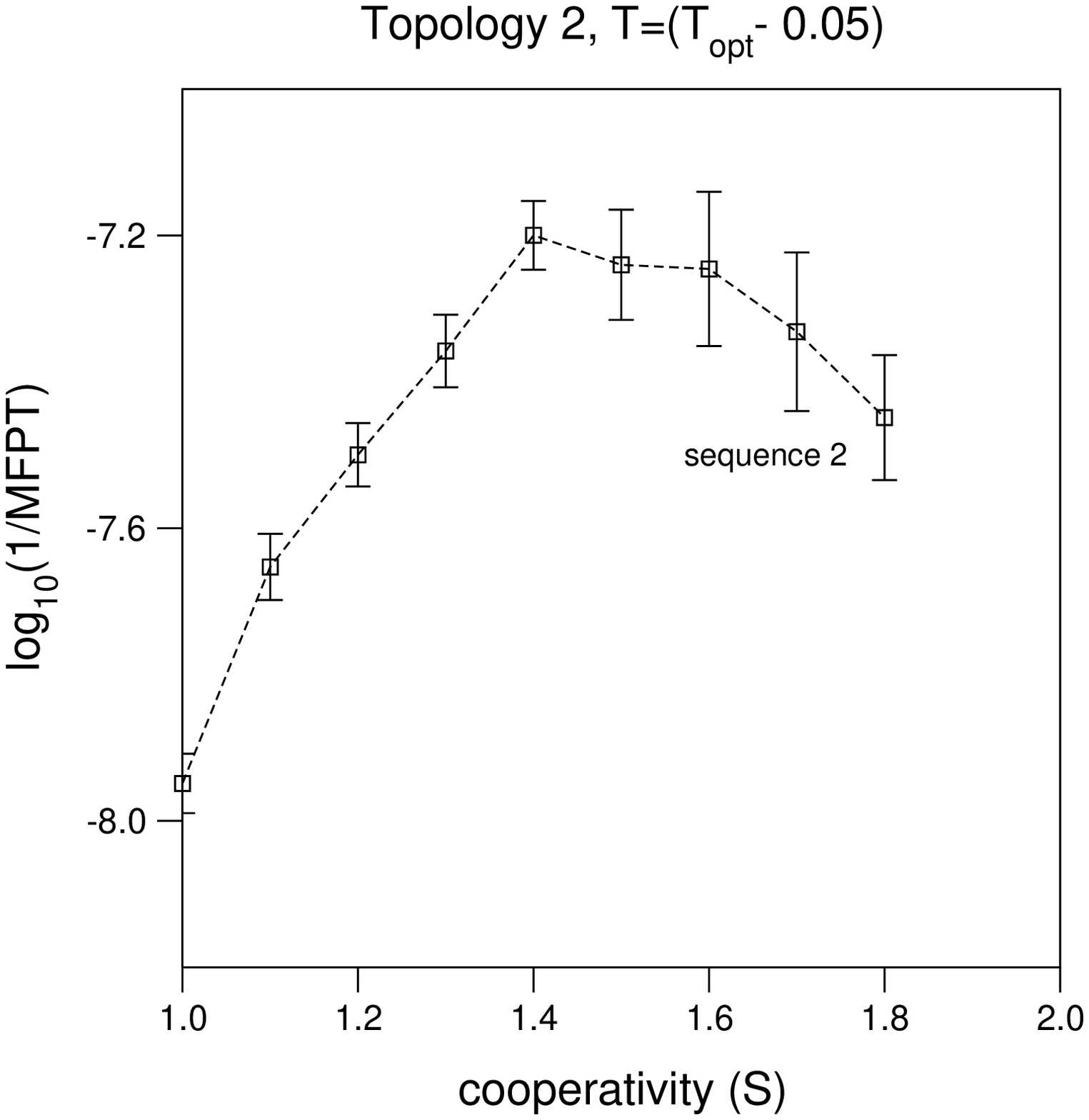}}}}
\caption{At the still lower temperature of T = T$_{opt}$ - 0.05 (corresponding to a $\sim 16^{\circ}$C  temperature drop for a protein with a T$_{opt}$ $\sim 40^{\circ}$C Ðe.g. the similarly sized fynSH3 domain51) the accelerating effects of enhanced cooperativity are even more pronounced. Under these conditions the folding rate of a sequences encoding topology 1 (top) increase by 5.7 to 6.9-fold as $S$ increases from unity to  $S_{opt}$. This is more than twice the maximum enhancement observed at T$_{opt}$and 2 to 40\% greater than the enhancement observed at T = (T$_{opt}$ - 0.03).  For a single sequence adopting the more complex topology 2 (bottom) we observe a 4.4-fold acceleration at this lower temperature, which is 80\% greater than that observed at T$_{opt}$ and 28\% greater than that observed at T = (T$_{opt}$ - 0.03).}
\label{fig:no4}
\end{figure}

If cooperativity accelerates folding by smoothing the energy landscape, it might be expected to have a larger effect at temperatures below T$_{opt}$ , where landscape roughness plays an increasingly important role in defining folding kinetics (Gutin 1996; Gutin 1998a) (under these conditions there is a higher probability for the chain to get trapped in metastable states). This is perhaps all the more true for sequences which, like those we have employed, were designed using an algorithm ensures their energy landscapes are relatively smooth (Shakhnovich and Gutin 1993). Consistent with this expectation we find that the effects of induced cooperativity are indeed much more striking at temperatures below T$_{opt}$ : at T = (T$_{opt}$  - 0.03) we observe up to 6-fold increases in folding rate (Figure 3, right). The cooperativity-induced increase in folding rates is, in fact, so great at T = (T$_{opt}$  - 0.03) that under these conditions (at $S$ = $S_{opt}$ ) all six sequences fold more rapidly than they do at T$_{opt}$  (for any $S$).  This presumably occurs because T$_{opt}$  represents a compromise between the driving force behind folding (native state stability), which increases as T decreases, and the kinetic consequences of trapped states, which tend to slow folding when T is reduced. Because of this, lowering the temperature below T$_{opt}$  and raising $S$ to $S_{opt}$  may accelerate folding by increasing the driving force (Gutin 1996; Gutin 1998b; Shakhnovich 1994) without concomitantly stabilizing trapped states that would otherwise slow the process. Perhaps consistent with this argument, it has previously been noted that rapidly folding lattice polymers tend to be those associated with a large gap between the energy of the native state and that of all other maximally compact states (Sali et al., 1994), an effect that might also arise due to a relationship between folding kinetics and thermodynamic cooperativity (Chan 2004).  To date, however, it has not yet been shown whether such an energy gap leads to thermodynamic cooperativity. 
\par
At the still lower temperature of 0.05 energy-units below T$_{opt}$  (which corresponds to a $\sim 16^{\circ}$C  temperature drop for a protein with a T$_{opt}$ of $\sim 40^{\circ}$C; see, e.g., ref Plaxco 1998) folding is significantly slowed and thus the exhaustive simulation of all six of our test sequences becomes difficult (particularly for sequences folding to the more complex topology 2). Nevertheless, we have investigated all three topology 1 sequences and a representative topology 2 sequence (sequence 2) under these conditions and find that the cooperativity-induced acceleration continues to increase as the temperature is reduced. For example, at this much lower temperature the three sequences encoding topology 1 fold 5.7 to 6.9 times more rapidly at Sopt than at $S$ = 1 (Figure 4, left). These accelerations are more than twice those observed at T$_{opt}$ and 40 to 200\% greater than those observed at 
T = (T$_{opt}$ - 0.03). Similarly, the folding of the sequence adopting the more complex topology 2 increases 4.4-fold at this temperature (Figure 4, right), which is 1.8 times the acceleration observed at T$_{opt}$ and 28\% greater than that observed at T = (T$_{opt}$ - 0.03).  These accelerations are so significant that, even at this very low temperature (which, if for no other reason, should slow folding due to the increase in the barrier height relative to $k_{B}$T), the folding of the cooperativity-optimized heteropolymers is significantly faster than that observed at T$_{opt}$ in the absence of added cooperativity.
We note, however, that although it is typical of this class of lattice models (see e.g. Faisca 2002a, Faisca 2005) the dispersion of folding times we observe  is smaller than that observed for single domain proteins (which span a six order of magnitude range). The chains we have investigated, however, are homogeneous in both length and stability, which perhaps accounts for the relatively limited dispersion in their folding rates

\subsection{\it {Cooperativity enhances rates by smoothing the energy landscape}}
\begin{figure*}[!ht] 
{\rotatebox{0}{\resizebox{7.5cm}{7.5cm}{\includegraphics
{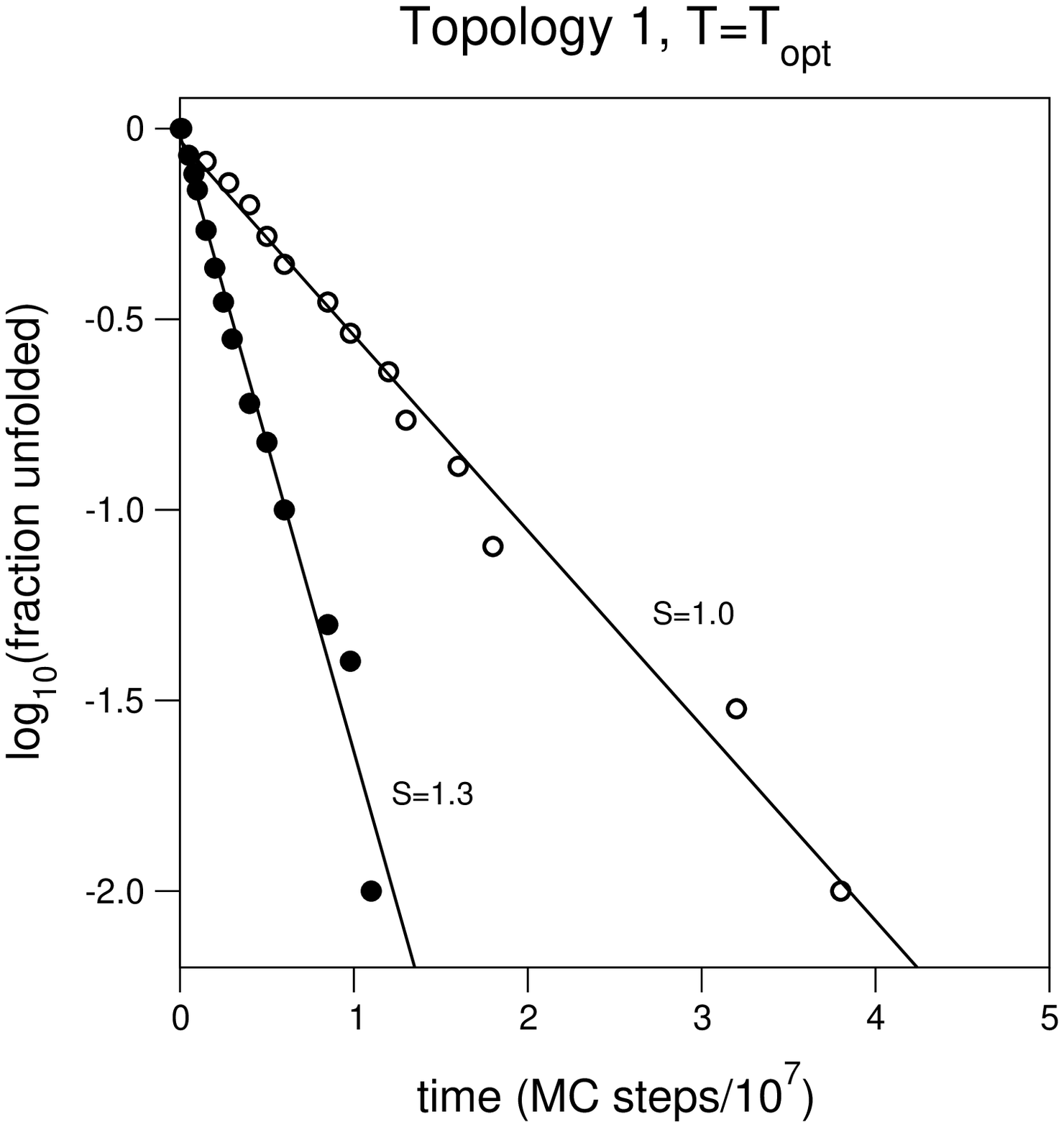}}}} \hspace{0.3cm}
{\rotatebox{0}{\resizebox{7.5cm}{7.5cm}{\includegraphics
{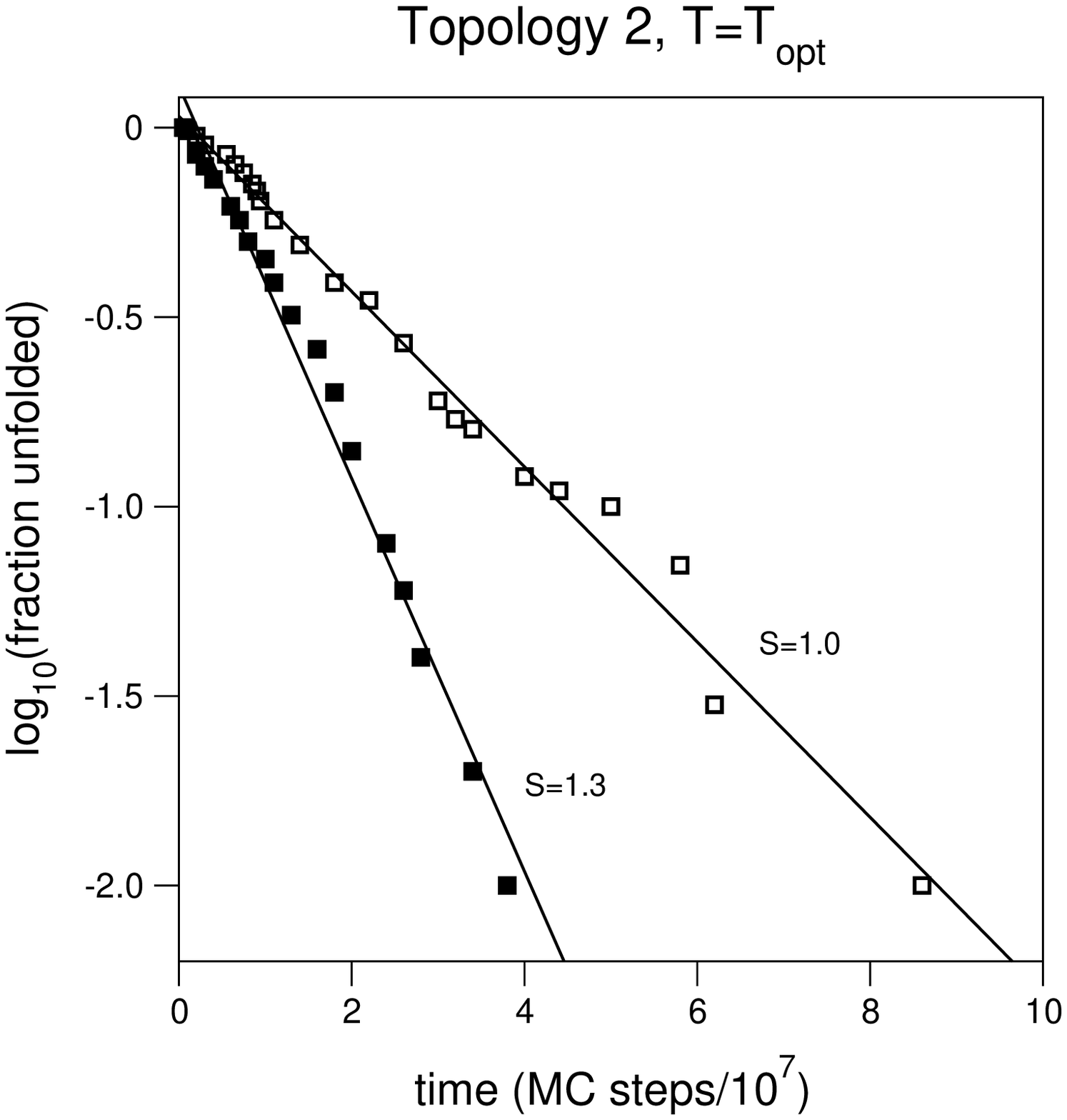}}}\vspace{0.8cm}} \\
{\rotatebox{0}{\resizebox{7.5cm}{7.5cm}{\includegraphics
{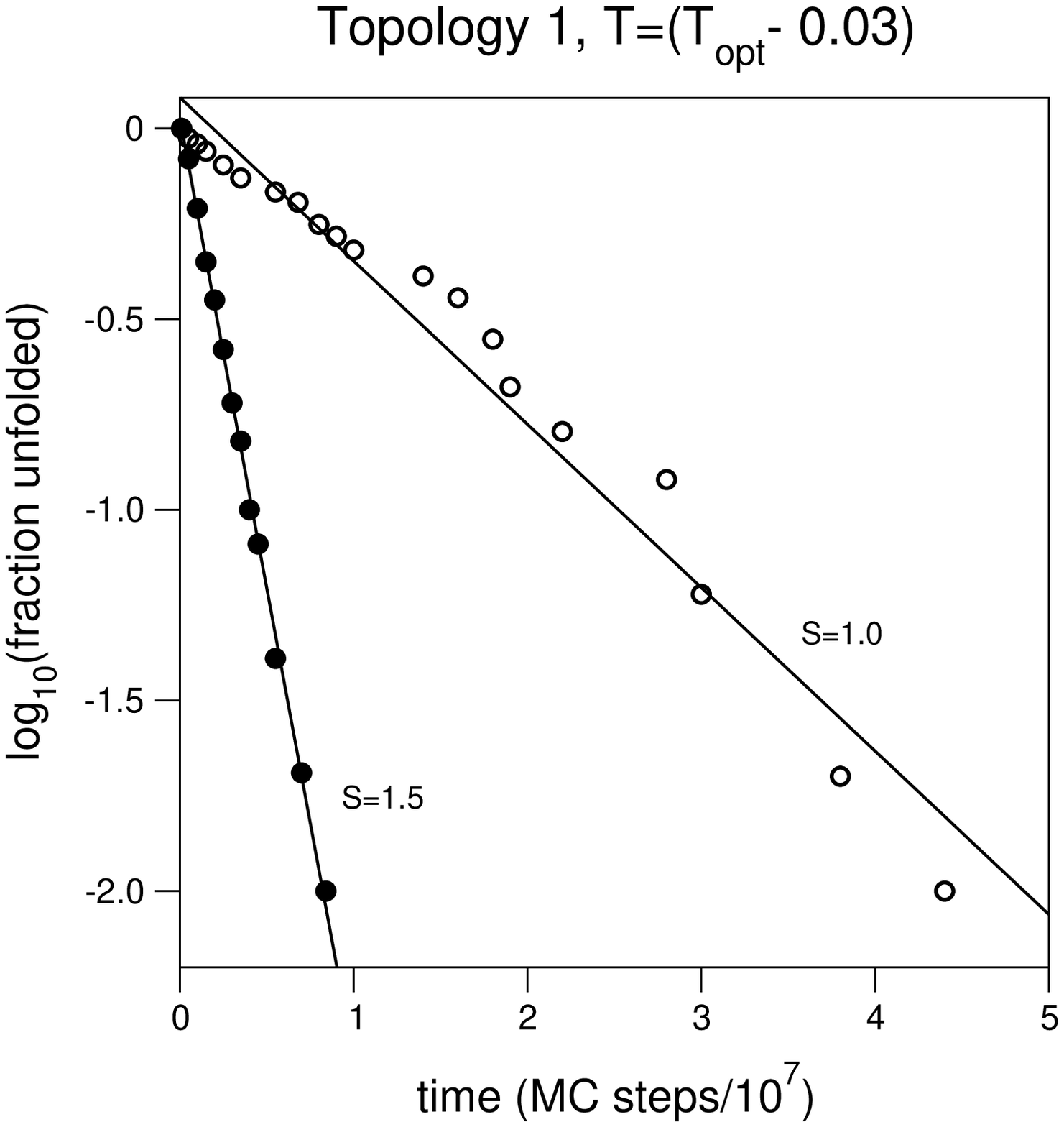}}}} \hspace{0.3cm}
{\rotatebox{0}{\resizebox{7.5cm}{7.5cm}{\includegraphics
{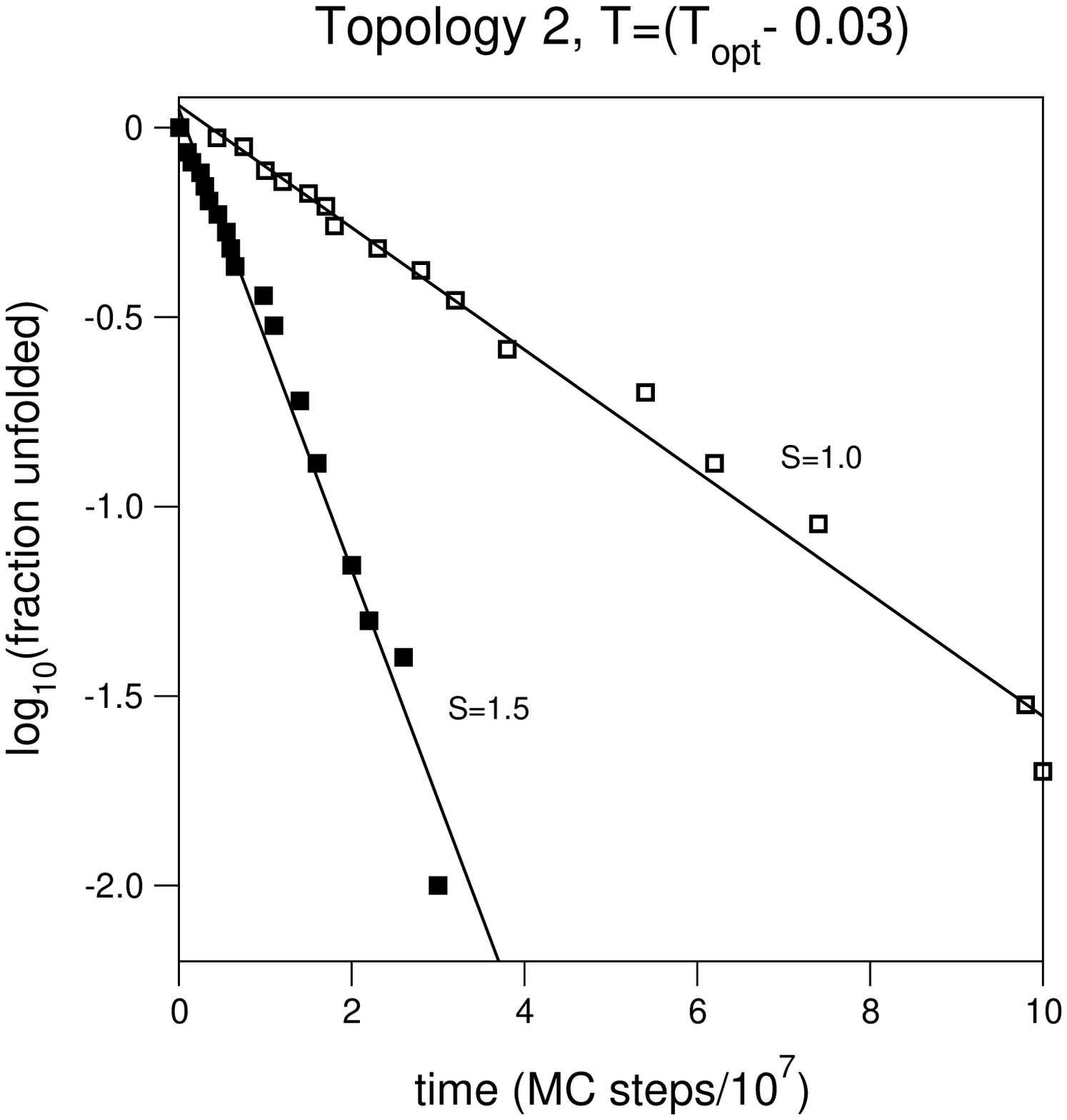}}}\vspace{0.8cm}} \\
{\rotatebox{0}{\resizebox{7.5cm}{7.5cm}{\includegraphics
{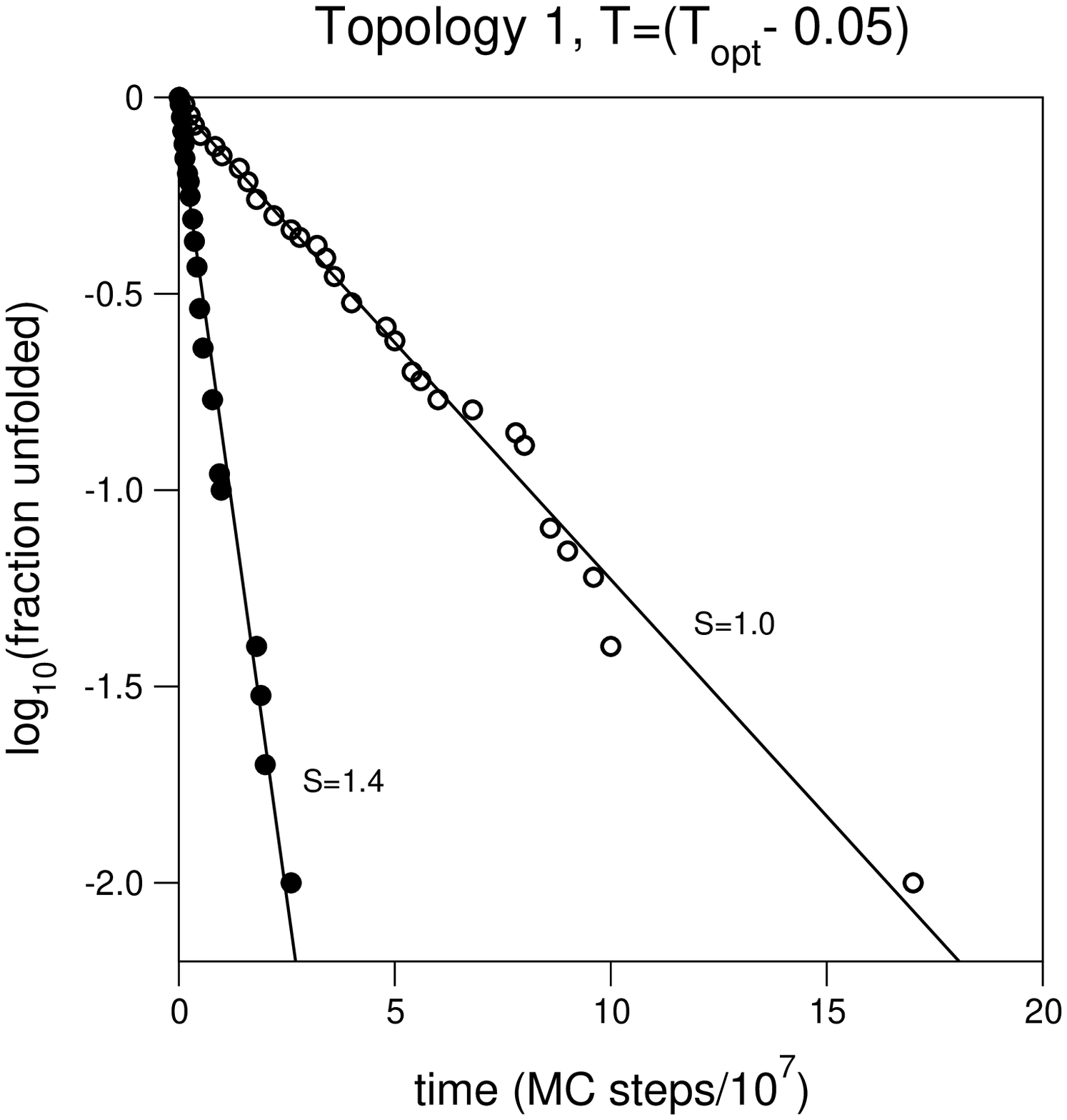}}}} \hspace{0.3cm}
{\rotatebox{0}{\resizebox{7.5cm}{7.5cm}{\includegraphics
{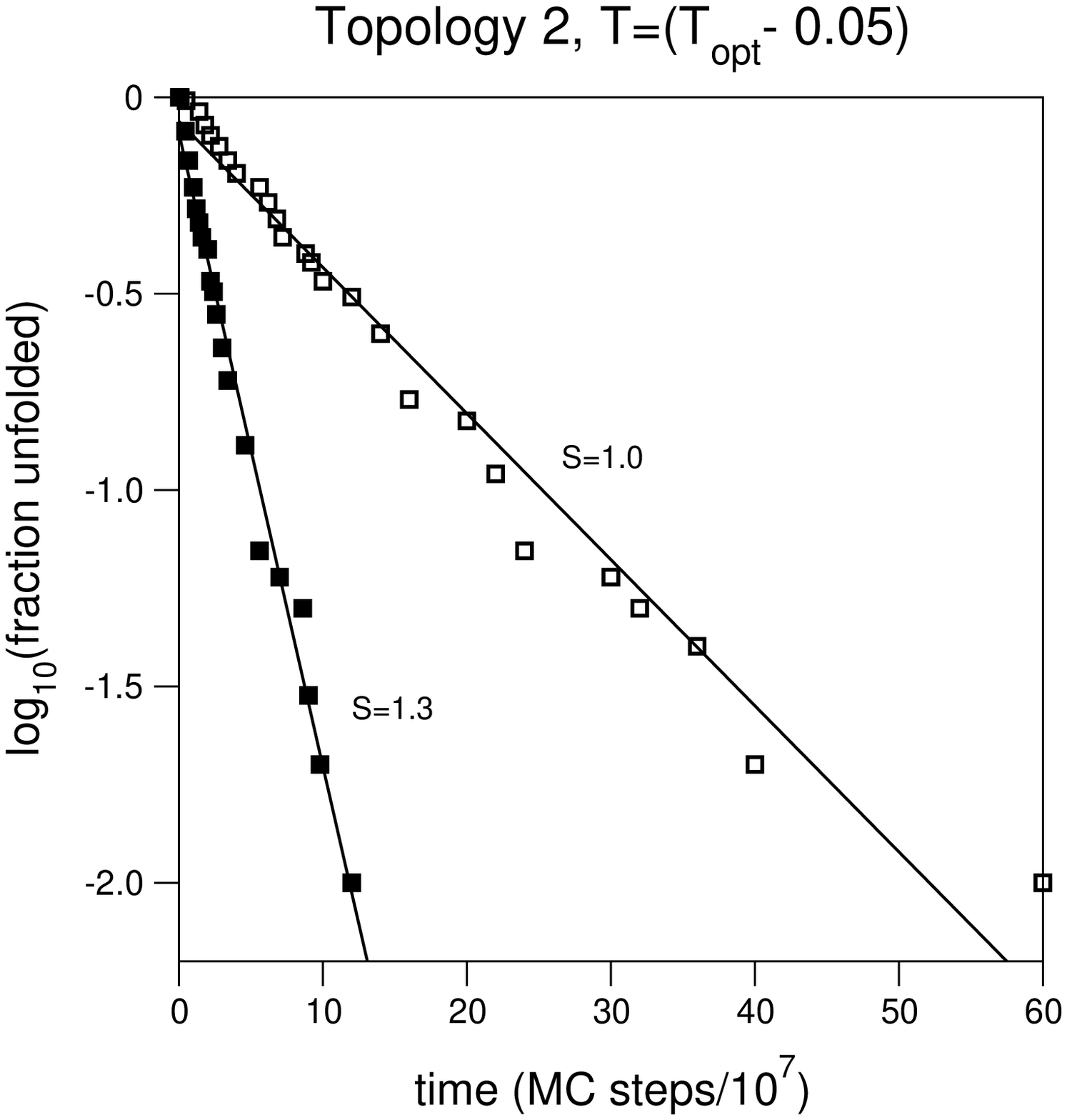}}}} \\
\caption{As the cooperativity increases to its optimal value ($S_{opt}$) the folding of MJ lattice polymers more closely approximates a single-exponential relaxation.   Shown are results for two representative sequences (sequence 1 of topology 1 and sequence 2 of topology 2) over a range of temperatures. While the folding of these sequences is generally well approximated as single exponential (because they were designed to fold efficiently), a distinct improvement in the single-exponential behavior of folding kinetics is nevertheless observed (from $R^{2}$ = 0.950 - 0.976 to $R^{2} >$ 0.988) when $S$ reaches $S_{opt}$.}
\label{fig:no5}
\end{figure*}

If thermodynamic cooperativity accelerates folding by reducing the stability of trapped states (i.e. smoothing the energy landscape), it might be expected to push kinetics from heterogeneous, non-exponential behavior towards simpler, single-exponential behavior (Bryngelson and Wolynes 1987; Onuchic 1997; Socci 1998; Nymeyer 1998). Consistent with this, Chan and Kaya have shown that significant thermodynamic cooperativity must be built into G\={o} -type lattice polymers in order to recover the linear chevron behavior observed for real proteins (Kaya and Chan 2003a).  Using two sequences (adopting topologies 1 and 2 respectively) we have investigated this question in more detail (Figures 5). We find that, while the folding of these sequences are generally rather well approximated as single exponential (because they were designed to exhibit smooth energy landscapes), without exception the kinetics of our models nevertheless move closer to single-exponential behavior as $S$ increases to its optimal value. For example, in the absence of added cooperativity we observe correlation coefficients (for the proportional relationship between $\log$(fraction unfolded) versus the number of MC steps) ranging from $R^2$ = 0.950 to 0.976. At $S_{opt}$, in contrast, these correlation coefficients invariably increase, ranging from 0.988 to 0.998.  Thus modest levels of added cooperativity appear to invariably improve the single-exponential behavior of even well designed sequences, suggesting that the added cooperativity is accelerating folding by smoothing the folding energy landscape.

\subsection{\it {Cooperativity may enhance the topology-dependence of folding rates}}

We (Jewett 2003), and others (e.g. Kaya and Chan 2003a; Ejtehadi 2004), have previously noted that modification of the G\={o}  potential to enhance its cooperativity enhances the extent to which G\={o}  polymer folding rates correlate with measures of native state topology. Given that we have investigated only two topologies in this study (albeit they are among the most and least topologically complex of the maximally compact 48-mers), we cannot similarly establish whether increasing cooperativity also enhances the topology-dependence of MJ polymer folding rates. Nevertheless, our results are consistent with this speculation: whereas, at T$_{opt}$, the three sequences that adopt the more complex (higher CO) topology 2 structure fold, on average, 6.2-times more slowly than those adopting to topology 1, this discrepancy almost doubles to 11.9-fold when $S$ rises to $S_{opt}$ (Figure 3). Whether this effect truly reflects a cooperativity-induced enhancement of the already significant topology-dependence of MJ polymer folding rates (Faisca and Ball 2002a), however, will require the simulation of significantly more topologies than were sampled here.

\section{\bf {Discussion}}

It is known that the folding of G\={o}  polymers, which are characterized by a native-centric energy potential and thus exhibit very smooth energy landscapes, slows as their cooperativity is increased (Jewett 2003; Kaya and Chan 2003a). Such increase in the folding time results from the fact that an increase in cooperativity leads, in such modified G\={o} -type models, to a destabilization of the transition state relative to the original G\={o}  model which, by considering only native interactions, is explicitly energetically biased towards the native state.  Here we have shown that, in contrast, the folding rates of energetically more complex -and thus perhaps more realistic- MJ models, where native as well as non-native interactions contribute to protein energetics, is increased when their thermodynamic cooperativity is increased and that this effect is enhanced as the temperature drops and landscape roughness plays a larger role in defining kinetics. The observed acceleration presumably arises because modest cooperativity increases folding rates more by destabilizing misfolded, trapped states than it decelerates the process by destabilizing the folding transition state (Eastwood and Wolynes 2001). In support of this claim, the folding kinetics of all six of the sequences we have investigated approach single-exponential behavior more closely as their cooperativity is increased and, as Chan and co-workers have noted (Kaya and Chan 2003a), increases in thermodynamic cooperativity tend to enhance linear chevron behavior.
\par
While our rescaling of the energy leads to a smoother energy landscape we should note that this does not necessarily imply that the potential we have employed becomes more G\={o}-like as $S$ is increased. G\={o} and G\={o}-like models exhibit smooth energy landscapes because they are native-centric, i.e., they neglect the effect of non-native interactions and only consider the contribution of native interactions to the protein's energy. The cooperativity term we have introduced, in contrast, enhances the stability of native-like conformations only when folding is near completion (i.e., for $Q > $0.7).  Moreover, this effect is mild at the modest levels of cooperativity employed here ($S< $1.5) .  For smaller values of $Q$ our rescaling procedure destabilizes the energy of any conformation with a certain fraction $Q$ of native contacts irrespective if that conformation is en-route to folding or not. This means that the stability of kinetic traps is not necessarily and selectively reduced relative to that of native-like conformations.  Indeed, would this be the case and, contrary to finding an optimal value for $S$, one would expect to observe a monatonic increase in folding rate with increasing $S$. The lack of such a relationship implies that the mechanism by which landscape smoothness is achieved in our model is differs from the mechanisms underlying the smoothness of G\={o} models and that, in turn, the observed accelerations are not a trivial outcome of the manner in which cooperativity has been encoded.
\par
The extent to which cooperativity accelerates folding appears to depend on both sequence-specific and topological effects. Indeed, it appears that increasing cooperativity may increase the spread in folding rates between simple and more complex topologies (Kaya and Chan 2003a; Jewett 2003; Ejtehadi 2004; Kaya 2005) in a manner reminiscent of the topology dependence of protein folding rates (Plaxco 1998a). This topology dependence presumably arises because, while the added cooperativity destabilizes energetically trapped, misfolded states (i.e. reduces so-called ``energetic frustration"), it does not affect the entropic barriers that arise from the polymer properties of the chain, such as connectivity (Makarov and Plaxco 2003) and excluded volume effects, and quirks of the native topology, such as lack of symmetry (Nelson 1997) (together sometimes termed ``topological frustration" (Shea 1999)). Such a linkage between folding rates, topology and cooperativity is consistent with the topomer search model (Makarov and Plaxco 2003), which postulates that two-state folding kinetics are defined by the diffusive search for unfolded conformations sharing the native topology (i.e. are in the ``native topomer"). If folding is thermodynamically cooperative, only the native topomer can form sufficient native-like interactions to ensure productive folding and thus the entropic cost of finding the native topomer will dominate relative folding rates. Similarly, while lattice polymers cannot fold via a strictly topomer-search process (a coarse lattice polymer cannot be in the native topology without actually being in the native state), the simulations described here illustrate the more general observation that, as cooperativity is increased, global properties such as native-state topology will play an increasingly important role in defining the folding barrier (Eastwood and Wolynes 2001; Plotkin 1997).
\par
Irrespective of the origins of the effect, previous studies clearly demonstrate that increasing thermodynamic cooperativity significantly enhances the topology-dependence of lattice polymer folding rates (Kaya and Chan 2003a; Jewett 2003; Ejtehadi 2004). Here we have shown that the addition of cooperativity also accelerates the folding of energetically complex lattice polymers and enhances the single-exponential nature of their kinetics. Taken together, these observations suggest that the observed topological dependence of two-state protein folding rates may be a consequence of the cooperativity necessary to ensure the smooth energy landscapes upon which rapid, single-exponential kinetics can occur.


\section{\bf {Methods}}

\subsection{\it The lattice polymer model}
We consider a simple three-dimensional lattice model of a protein molecule with chain length $L$ = 48.  Individual amino acids are represented in this model by beads of uniform size occupying lattice vertices. These monomer units are connected into a polypeptide chain using bonds of uniform (unitary) length corresponding to the lattice spacing. 
\par
\begin{figure}
{\rotatebox{0}{\resizebox{8.0cm}{8.0cm}{\includegraphics{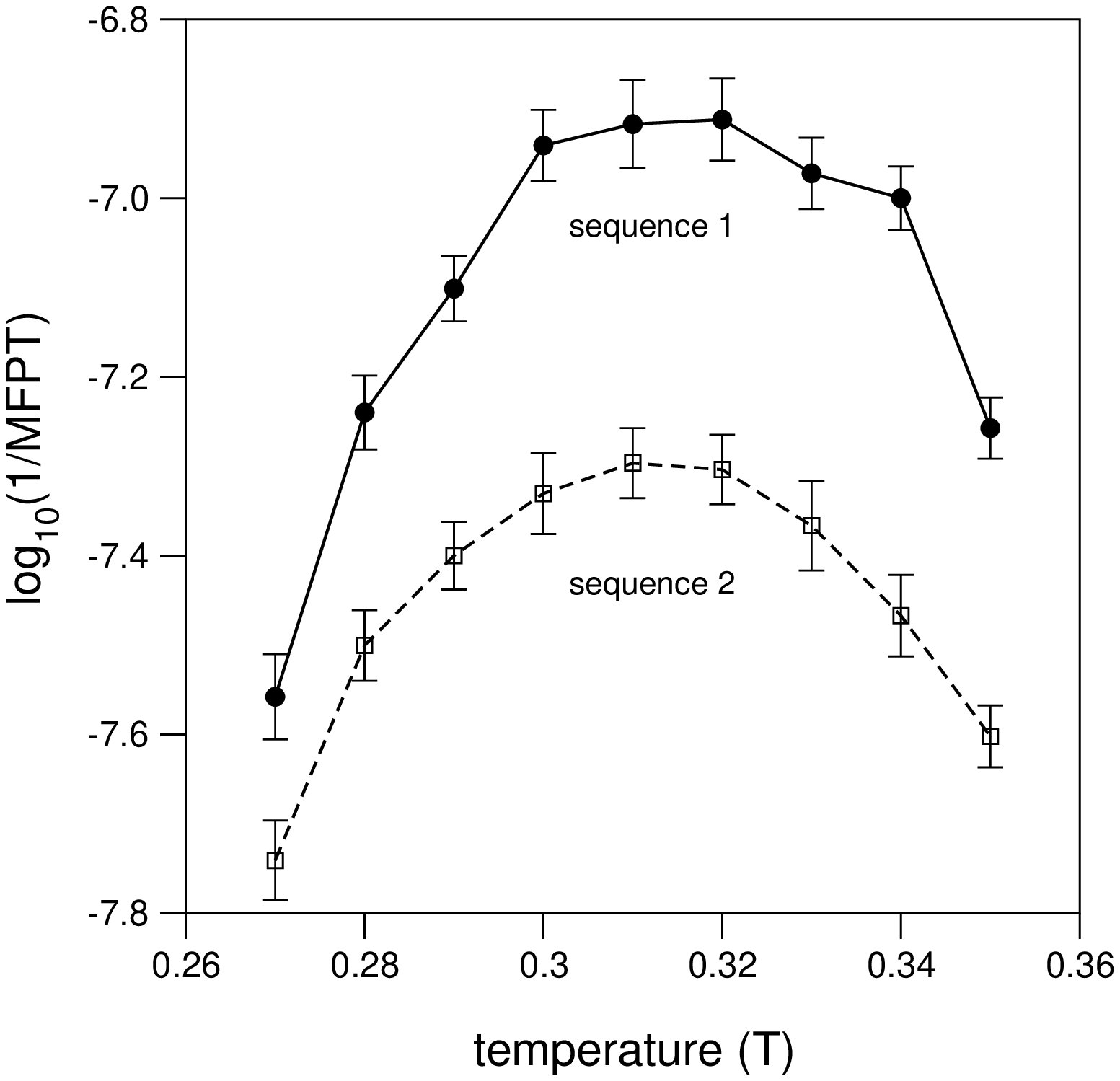}}}}
\caption{ The folding kinetics of MJ lattice polymers are strongly temperature dependent. Shown is the dependence of the logarithmic folding rate, $log_{10}(1/MFPT)$, on the simulation temperature T, for sequence 1 of topology 1 and sequence 2 of topology 2, under conditions of no induced cooperativity ($S$ = 1). The optimal folding temperature is the temperature at which folding is most rapid.}
\label{fig:no6}
\end{figure}
In order to mimic folding to the native state we use a standard Monte Carlo (MC) algorithm together with the kink-jump move set (see e.g. Landau and Binder 2000) and with the energy defined as in Equation 2. Accordingly, random local displacements of one or two beads are repeatedly accepted or rejected in accordance with the standard Metropolis rule (Metropolis 1953).  All MC simulations were started from randomly generated unfolded conformations and their folding dynamics were traced by following the evolution of the fraction of native contacts, $Q = n/N$, where $N$ is the total number of native contacts and $n$ is the number of native contacts formed at a given MC step.  Each simulation was run until the polymer reached the native fold, i.e. when $Q$ = 1. The number of MC steps required to fold to the native state was defined as the first passage time and the folding time was computed as the mean first passage time (MFPT) of 100 simulations.

\subsection{\it Structure selection and sequence design}

\begin{table*}
\caption{\label{tab:no1} Geometric traits for native topologies 1 and 2.
Contact order, CO, number of long-range contacts $Q_{D}$, we define as long-range (LR) a contact between any two contacting beads $i$ and $j$ whose sequence separation is at least 12 units, i.e. $|i-j| \geq 12$  (Gromiha and Selvaraj 2001; Makarov and Plaxco 2003).
Long-range order, LRO, which is equivalent to $Q_{D}$ normalized by chain length (Gromiha and Selvaraj 2001)}
\begin{ruledtabular}
\begin{tabular}{c c c c}
Topology \#     &CO&   $Q_{D}$  &  LRO  \\ 
\hline 
1 &    $0.13$ &     $0.33$  & $0.40$  \\
2 &    $0.26$ &     $0.74$  & $0.88$  \\
\end{tabular}
\end{ruledtabular}
\end{table*}

\begin{table*}
\caption{\label{tab:no2}Designed sequences. The folding rate is measured 
at the optimal folding temperature T$_{opt}$.}
\begin{ruledtabular}
\begin{tabular}{c c c c c c}
\# (Topology \#) & Sequence      & E$_{nat}$&       T$_{opt}$  & T$_{f}$ & $-\log_{10}(t)$  \\ 
\hline
  $1$ (1) & \texttt{FRTRPLNHDFYNYKIWEPFKPADFPKAWDRMLDHVWDSMASWGHQHCS}   
&      $-25.85$  &      0.32  & 0.32  &   -6.91 $\pm$ 0.05 \\
  $2$ (1) & \texttt{CDLPPFTYRHHGNDFWKNYEMIKHWDLWRDMFRAFWSDPVKASPHQAS} 
&      $-25.92$  &      0.32  & 0.32  &   -6.42 $\pm$ 0.03 \\
  $3$ (1) & \texttt{FRTPWVSHQFYAYKLMEHFKWGDFCRNMDKWIDSLPDRWNPAPHDHAS}
&      $-26.09$  &      0.32  & 0.32  &   -6.18 $\pm$ 0.04  \\
  $1$ (2) & \texttt{KDKIHFRMNYGYPAWDAQSVKDLTCPRDWHFPHMRDPSHNWELAFFWS} 
&      $-25.87$  &      0.34  & 0.33  &   -7.47 $\pm$ 0.05 \\
  $2$ (2) & \texttt{ENDVTMDMDPSPCLFRIHNLPRAHSFDRFGWHQFDKYHYKWKWAWAPS} 
&      $-26.15$  &      0.31  & 0.34  &   -7.30 $\pm$ 0.04 \\
  $3$ (2) & \texttt{EHDAQLDFDWSRWTWHGRNSYHAPAMYRWPVHDMDKPNPKFKIFFLCS}        
&      $-26.24$  &      0.32  &0.33   &   -7.44 $\pm$ 0.06  
\end{tabular}
\end{ruledtabular}
\end{table*}

We selected two model topologies from a pool of ~500 maximally compact conformations (MCC) found using simulations of homopolymer relaxation.  The latter are MC simulations where a polymeric chain composed by beads of the same chemical species is launched, at temperature T = 0.7, from a randomly generated conformation and relaxes to the minimum energy conformation. For self-attractive homopolymers of length $L = 48$ on a three-dimensional cubic lattice, the most stable conformation is a MCC cuboid with 57 contacts.  We have computed a histogram of the contact order frequencies of a large sample of these MCCs (data not shown) and identified the two extremes at CO values of 13\% and 26\%.  These topologies, numbered here 1 and 2 respectively, were selected for our studies (Figure 1 and table 1).
\par
Ensembles sequences were designed to fold into topologies 1 and 2 (table 2).  The goal of the design process was to identify sequences that fold rapidly and efficiently into a pre-selected conformation named target structure. We followed Shakhnovich and Gutin (Shakhnovich and Gutin 1993) and Shakhnovich (Shakhnovich 1994) in attempting this by seeking the sequence with the lowest possible energy in the target state, as given by Equation 1.  To this end the target's coordinates were quenched and the energy of Equation 1 annealed with respect to the sequence variables $\sigma$.  This amounts to a simulated annealing in sequence space: starting from some randomly generated sequence, transitions between different sequences (which are generated by randomly permuting pairs of beads) are successively attempted along with a suitable annealing schedule.

\subsection{\it Optimal folding temperature}

The existence of optimal folding temperature (T$_{opt}$), corresponding to the temperature at which the folding rate is maximized, has extensively reported in lattice studies of protein folding (Gutin 1996; Gutin 1998b; Shakhnovich 1994; Faisca and Ball 2002; Cieplak 1999; Faisca and Ball 2002a; Faisca 2005). In order to determine T$_{opt}$ we have computed the folding times (in the absence of added cooperativity) over a broad temperature range for all six of the sequences we have investigated. Figure 6 illustrates the dependence of the logarithmic folding rates, $log_{10}(1/MFPT)$, on the simulation temperature for sequences 1 and 2 folding to topologies 1 and 2 respectively. These sequences exhibit T$_{opt}$ of 0.32 and 0.31 respectively. Below and above these optimal temperatures their folding slows significantly, with the stronger temperature dependence being observed at lower temperatures. 
\par
For comparison with prior lattice polymer folding studies we have determined the folding transition temperatures, T$_{f}$, of the six sequences we have employed.  T$_{f}$ -typically denoted as the melting temperature or T$_{m}$ in the experimental literature) is the temperature at which denatured states and the native state are equally populated at equilibrium.  In the context of a lattice model it can be defined as temperature at which the average value $<Q>$ of the fraction of native contacts is equal to 0.5 (Abkevich 1995). In order to determine T$_{f}$ we averaged $Q$, after collapse to the native state, over MC simulations lasting at least 20 times longer than the average folding time computed at T$_{opt}$,.  Similarly, in order to compute population histograms at T$_{f}$ as a measure of thermodynamic cooperativity (Figure 2) data were averaged after collapse to the native state over MC simulations least 20 times longer than the average folding time at T$_{opt}$.


\section{\bf {Acknowledgements}}PFNF would like to thank Funda\c c\~ao para a Ci\^encia e Tecnologia for financial support through grants SFRH/BPD/21492/2005 and POCI/QUI/58482/2004. This work was also supported by NIH grant R01GM62868-01A2 (to KWP). The authors wish to thank Andrew Jewett, Vijay Pande and Joan-Emma Shea for providing critical commentary and Hue Sun Chan for helpful discussions.

\end{document}